\documentclass[pre,aps,twocolumn,floats,showpacs,amsmath,amssymb]{revtex4-1}
\usepackage{amsmath,amssymb}
\usepackage{bbm}
\usepackage{graphicx}

\usepackage{xcolor}

\newcommand{\fref}[1]{Fig.~\ref{fig:#1}}

\newcommand{\flabel}[1]{\label{fig:#1}}
\newcommand{\eref}[1]{Eq.~\ref{eqn:#1}}

\newcommand{\elabel}[1]{\label{eqn:#1}}

\newcommand{\beq}{\begin{equation}}
\newcommand{\eeq}{\end{equation}}
\newcommand{\XP}{\mathbf{x}}       
\newcommand{\PP}{\mathcal{P}}      
\newcommand{\DD}{\mathcal{D}}      

\begin{document}
\title{
A method of incorporating rate constants as kinetic constraints \\in molecular dynamics simulations} 

\author{Z. Faidon Brotzakis$^\dagger$, Michele Vendruscolo$^\dagger$ and Peter G. Bolhuis$^\ddagger$}

\affiliation{$\dagger$  Department of Chemistry, University of Cambridge, Cambridge CB2 1EW, UK.\\$\ddagger$ van 't Hoff Institute for Molecular Sciences, University of Amsterdam,  PO Box 94157, 1090 GD Amsterdam, The Netherlands}
\date{\today}

\begin{abstract}
From the point of view of statistical mechanics, a full characterisation of a molecular system requires the experimental determination of its possible states, their populations and the respective interconversion rates. Well-established methods can incorporate in molecular dynamics simulations experimental information about states using structural restraints, and about populations using thermodynamic restraints. However, it is still unclear how to include experimental knowledge of interconversion rates.
Here we introduce a method of imposing known rate constants as constraints in molecular dynamics simulations, which is based on a combination of the  maximum entropy and maximum caliber principles. Starting from an existing ensemble of trajectories, obtained from either molecular dynamics or enhanced trajectory sampling, this method provides a minimally perturbed path distribution consistent with the kinetic constraints, as well as a modified free energy and committor landscape. 
We illustrate the application of the method to simple toy systems, as well as to all atom molecular simulations of peptide association and folding. We find that by combining experimental rate coefficient data and molecular dynamics simulations we are able to determine new transition states, reaction mechanisms and free energies. For instance, in the case of chignolin protein folding we find that imposing a slower folding rate shifts the transition state to more native like conformations, while it increases the stability of the unfolded region. We foresee this method can extend the applicability of both atomistic and coarse-grained molecular simulations as an accurate kinetic tool in structural biology as well as
assist amending imperfections in current atomistic force fields to reproduce the kinetics and thermodynamic observables. Finally, the approach is general, and applicable to a wide range of systems in biology, physics, chemistry, and material science.  

\end{abstract}
\maketitle

\section{Introduction}
The first step in the study of a molecular system typically consists in the determination of its conformation, as for example most commonly done by using X-ray crystallography (X-ray), cryo-electron microscopy (cryo-EM) or nuclear magnetic resonance spectroscopy (NMR) for obtaining the structures of proteins and of nucleic acids~\cite{Alberts2002}. By revealing a wide range of structure-function relationships, this approach has enabled major advances in molecular  biology~\cite{Alberts2002}. From a procedural point of view, experimental measurements, such as electron densities in X-ray and cryo-EM or interproton distances in NMR, combined with well-established theoretical chemistry rules,  facilitates the building of molecular structures using computational methods~\cite{Brunger1998}.

As at the molecular level under physiological conditions thermal fluctuations are relevant, it is becoming increasingly common to perform a second step, which involves the determination of the structures of the thermally excited states of a molecular system, together with their populations~\cite{Mittermaier2006}\footnote{'excited' refers to the dynamical structures that the system visits rather than electronic excitation}. This goal is typically achieved by incorporating experimental measurements as structural restraints in molecular dynamics simulations to sample the free energy landscape~\cite{Bonomi2017}. The maximum entropy principle (MaxEnt) provides a rigorous framework to implement this strategy.  To carry out this step,  a range of methods are now available~\cite{Cavalli2013a,Boomsma2014}, 
resulting in the determination of a 'thermodynamic ensemble' of structures~\cite{Bonomi2019}.

One may not, however, stop at this level if kinetic properties are to be characterized. As a third step in the determination of a molecular system, one would like to obtain a 'kinetic ensemble', comprising the structures of the different states of a molecular system, their populations and their interconversion rates~\cite{Bonomi2019}. Approaches for determining kinetic ensembles are not readily available, as there is no well-established method of incorporating experimental information about kinetic rates in molecular modeling procedures. Our aim here is to make a first step in this direction.

To achieve this goal, we start, as  seems quite natural, from a MaxEnt approach, where one maximizes a configurational entropy, subject
to constraints given by experimental data, in order to predict a new
configurational probability distribution. MaxEnt can also model  uncertainties in the data effectively turning constraints into restraints~\cite{Cavalli2013a,Boomsma2014}.  Addressing the problem in various ways as a Bayesian or a maximum likelihood problem~\cite{Cesari2018,Boomsma2014,Pitera2012,Hummer2015, Bonomi2015, Olsson2017}, leads  to numerous applications for example in cases where force fields are less accurate, such as for intrinsically disordered proteins (IDPs) and
RNA~\cite{Heller2017,Heller729392,Borkar2016,Cesari2018,Orioli2019}. Apart from ensemble refinement, application of MaxEnt yields perturbative correction terms to the potential energy along order parameters or collective variables relating to the experimental data~\cite{Cesari2016a,Bonomi2017,Cesari2018}. 

To enforce experimental information about rate constants the MaxEnt method can be combined with the maximum caliber principle (MaxCal)~\cite{Jaynes1980}. This approach seems again quite natural, as MaxCal is a general variational framework of non-equilibrium and equilibrium statistical mechanics
 with a wide scope, from flux-fluctuation relationships to pathway distributions~\cite{Jaynes1980,Monthus2011}. In MaxCal, one maximizes a path entropy over all possible pathways, subject to dynamical constraints such as average fluxes, in order to predict relative path weights~\cite{Presse2013}. Rigorous and general MaxCal implementations have found so far fewer applications compared to MaxEnt approaches due to the  difficulty both in sampling  path distributions of  complex systems and in acquiring experimental data about rate constants. For example  MaxCal enabled
 reweighting of the equilibrium distribution of macrostates given experimental rate constants for  
 Markov State Models (MSM) or time discrete pathways~\cite{Dixit2018,Filyukov1967}. A recent implementation of the MaxCal for time-resolved data imposed  time-dependent constraints along a few degrees of freedom of the system, 
 or collective variables (CV),  to agree with time-resolved experimental
 data~\cite{Capelli2018}. MaxCal methods have also been extended to
 non-equilibrium dynamics~\cite{Presse2013,Bereau2020}. 
An important aspect, however, is that such methods rely on spatially discrete models, on limited time resolved data, or on biased dynamics, while in practice one usually only has access to experimental rate or diffusion constants. As starkly captured by Jaynes, ``reconstructing MaxCal path ensembles containing the microscopic space and time dynamics is  difficult''~\cite{Jaynes1985}.  While theoretically rigorous, the MaxCal formalism has not been implemented to date for reweighting purposes in time and space continuous unbiased trajectories. 

Here we present a method of determining kinetic ensembles using the MaxCal strategy, by reweighting  path ensemble distributions a posteriori, according to both kinetic and thermodynamic experimental data. The methodology yields experimentally-corrected free energy and committor landscapes, and provides structural ensembles that exhibit accurate configurations, including in the regions of the barrier between states. 
 
Given experimental forward and backward rate constant constraints we combine MaxEnt with MaxCal to find a biasing function that simultaneously acts on equilibrium and rate constants. This bias function gives  correcting weights to the pathways of the  equilibrium path ensemble distribution. The  equilibrium path ensemble distribution is generated from computing reweighted path ensembles (RPEs)~\cite{Rogal2010a} based on either long molecular dynamics trajectories, or on enhanced sampling of trajectories, e.g., using Transition Interface Sampling (TIS) simulations~\cite{vanErp2003}, or in one step using the Virtual Interface Exchange Transition Path Sampling (VIE-TPS) method~\cite{Brotzakis2019e} for pathways sampled by Transition Path Sampling (TPS)~\cite{Dellago_1998}. Such TPS based methods focus on reactive or partially reactive pathways, thereby bypassing the computationally expensive sampling of the stable states. In the remainder of the text, RPE will refer to the reconstructed equilibrium (reweighted) path ensemble distribution from simulation, while ``kinetic ensemble'' will refer to the equilibrium path ensemble distribution after imposing the experimental  kinetic constraints.

While our  method  applies constraints to the distributions, uncertainties in the experimental and simulations data can be taken into account\cite{Bonomi2015}. 
While constraints impose strict equality with experimental data, restraints impose equality within errors in the data. MaxEnt and MaxCal allow to model uncertainties in the data, effectively turning constraints into restraints

In this work we focus on biological problems without losing the generality of our statements. Thus, our approach can be applied to all molecular dynamics  simulations where trajectory reweighting to match target kinetics is possible and helpful.   

This paper is organised as follows. In Section~\ref{sec:theory} we introduce the background theory and new concepts of our approach. In Section~\ref{sec:results} we illustrate the approach, and we end with an outlook in  Section~\ref{sec:conclusions}.

 \section{Theory}
\label{sec:theory} 

\subsection{MaxEnt in configuration space}

In this section we briefly recapitulate how the MaxEnt can be used to combine simulations with  experimental data~\cite{Cavalli2013a,Boomsma2014}. 
In its original formulation, MaxEnt states that the probability distribution of the states of a system maximally compatible with a set of observed data is the one maximizing the associated Shannon entropy. This principle has been extended to a maximum relative entropy principle, which has the advantage of being invariant with respect to changes of coordinates and coarse-graining, and has been shown to play an important role in multiscale problems~\cite{Orioli2019}. The entropy is here computed relative to a given prior distribution $P^0(x)$ and, for a system described by a set of continuous variables $x$, e.g. the positions and velocities of all atoms in a molecular system,  is defined as
\begin{equation}
\elabel{eq:RelEnt}
 S[P||P^0] = -\int dx P(x) \ln \frac{ P(x)}{P^0(x)}.
\end{equation}
This entropy can be maximized as 
\begin{align}
  P^{ME}(x) =&\operatorname*{argmax}_{P(x)}  S[P||P^0]  \\    
&\textrm{subject to:}  \begin{cases}
\int dx P(x) s_i (x) = \langle s_i (x) \rangle =s_i^{exp} \\
  \int dx P(x)=1
\end{cases} \notag
\end{align}
where 
the experimental observations
$s_i^{exp}$ ($i \in \{  1, 2 \dots M\} $) constrain the ensemble average of $M$ observables s$_i(x)$, computed over the distribution $P(x$), to be equal to s$_i^{exp}$, and an additional constraint ensures that the distribution $P(x)$ is normalized. $P^0$(x)  is called the ``prior'' probability distribution, encoding the knowledge available before the experimental measurement. $P^{ME} (x)$ instead represents the best estimate for the probability distribution after the experimental constraints have been enforced, and is thus called the ``posterior'' probability distribution. The subscript ME denotes the fact that this is the distribution that maximizes the entropy.

Since the relative entropy $S[P||P^0]$ is the negative of the Kullback-Leibler (KL) divergence $D_{KL}[P||P^0$] , the procedure described above can be interpreted as a search for the posterior distribution that is as close as possible to the prior knowledge and agrees with the given experimental observations. In terms of information theory, the KL divergence measures how much information is lost when prior knowledge $P^0$(x) is replaced with $P(x)$. Always non-negative, the KL divergence is a measure of the difference between the distributions, and vanishes only if the two distributions are identical. 

A powerful approach to solve the maximization problem in ~\eref{eq:RelEnt} is based on the method of Lagrange multipliers, namely searching for the stationary point of the following Lagrange function:
\begin{align}
\elabel{eq:LagrEnt}
\mathcal{L} & = S[P||P^0]-  \sum_{i=1}^{M}  \mu_i \left ( \int dx s_i
  (x) P(x) - s_i^{exp}\right ) \notag \\
 &- \nu \left ( \int dx P(x) - 1\right ), 
\end{align}
where $\mu_i$ and $\nu$ are suitable Lagrange multipliers taking care of the experimental observations and the probability normalization, respectively. The functional derivative of $\mathcal{L}$ with respect to  $P(x)$  is
\begin{equation}
\elabel{eq:LagrEnt_der}
\frac{\delta \mathcal{L} }{\delta P(x)}= -\ln \frac{ P(x)}{ P^0(x)} -1  -\sum_{i=1}^{M}  \mu_i   s_i (x)  - \nu. 
\end{equation}

By setting $\frac{\delta \mathcal{L} }{\delta P(x)}=0$ and neglecting the normalization factor, the posterior reads
\begin{equation}
\elabel{eq:PostEnt}
P^{ME}(x) \propto e^{-\sum_{i=1}^{M}  \mu_i   s_i(x) } P^{0}(x)
\end{equation}
%

Solving \eref{eq:LagrEnt_der} turns out to be equivalent to minimizing the  function
\begin{equation}
\elabel{eq:Gamma}
\Gamma ({\boldsymbol \mu})= \ln \left [   \int  dx e^{-\sum_{i=1}^{M}  \mu_i   s_i(x) } P^{0} (x)  \right ] + {\boldsymbol \mu} \cdot {\boldsymbol s^{exp}}
\end{equation}
with respect to $\mu_i$, leading to the equation(s) $\langle s_i \rangle = \int dx s_i
  (x) P(x) =s_i^{exp}$, and thus giving for each observable the Lagrange multiplier $\mu_i$.

We also note that MaxEnt can  model uncertainties in the data, i.e  the experimental errors\cite{Bonomi2015,Cesari2018}. This is done by adding the expected error due to the perturbed distribution $\langle  e_i \rangle $ to the constraint average, i.e. $\langle s_i \rangle =  s^{exp}_i + \langle  e_i \rangle $.
For a Gaussian distributed error with a standard deviation $\sigma_i$ the average error is $\langle  e_i \rangle = - \mu_i \sigma_i^2 $, with $\sigma_i$ the level of confidence in the data, e.g. experimental measurements. 
Adding this to \eref{eq:Gamma} yields
\begin{equation}
\elabel{eq:Gammaerror}
\Gamma ({\boldsymbol \mu})= \ln \left [   \int  dx    P^{ME} (x)  \right ] + {\boldsymbol \mu} \cdot {\boldsymbol s^{exp}} +  \frac{1}{2}\sum_{i=1}^{M}\mu_i^2 \sigma_i^2. 
\end{equation}
Minimizing this function leads to a solution of the Lagrange multipliers $\mu_i$ that  account for the error. If $\sigma=0$ the situation is identical to \eref{eq:Gamma}, while if $\sigma$ is large the Lagrange multiplier will be close to zero, almost not perturbing the original distribution.
In this way the constraint on the distribution is turned into a restraint, depending on the level of confidence in the data. 
In most of our presentation, we will discuss imposing constraints, although one should keep in mind that it is always possible to extend the results to imposing restraints, using the above procedure.  

\subsection{MaxCal in path space}
The MaxEnt principle can be straightforwardly extended to trajectory space\cite{Presse2013}.  Consider a prior path probability distribution  $\PP^0[\XP]$ of trajectories  $\XP$, each consisting of $L$ frames  $\XP=\{ x_0, x_1,  \dots x_L  \}$, where subsequent frames are separated by a time interface $\Delta t$, such that the total duration of a path is $\mathcal{T} = L \Delta t$.
Here we assume that the path represents a dynamical evolution according to the equations of motion, as given e.g. by a MD simulation,  and contains reliable dynamic information, of course up to the extent of the resolution and faithfulness of the force field. 
The (relative) path entropy, or caliber, for any path distribution $\PP[\XP]$, is 
\begin{equation}
\elabel{eq:RelEntPath}
 S[\PP||\PP^0] = -\int \DD{\boldsymbol \XP} \PP[\XP] \ln \frac{ \PP[\XP]}{ \PP^0[\XP]},
\end{equation}
where $\DD \XP$ indicates an integral over all trajectories or paths $\XP$. 
  The  maximum caliber principle states that the optimal distribution $\PP^{MC}[{\XP}]$ is given by 
\begin{align}
  \PP^{MC}[{\XP}]=&\operatorname*{argmax}_{\PP[\XP]}  S[\PP||\PP^0]  \\    
&\textrm{subject to: }\begin{cases}
\int \DD{\XP }  \PP[{\XP }] s_i [{\XP }] = \langle s_i [{\XP}] \rangle =s_i^{exp} \\
  \int \DD{\XP } \PP[{\XP }]=1
\end{cases}\notag
\end{align}
that is, $\PP^{MC}[{\XP}]$  maximizes the path entropy or caliber, while obeying the constraints given by external constraint $s_i^{exp}$.
The observable ensemble average $\langle s_i [{\XP}]
\rangle$ can relate to any measurement either giving rise to
static/thermodynamic or dynamic/kinetic information.

Starting with dynamical  information, 
consider an arbitrary time correlation function 
\begin{equation}
c(t) = \langle s_i(0) s_j(t) \rangle = \int \DD{ \XP}
\PP[\XP]  s_i(x_0) s_j(x_\tau),
\end{equation} 
where $\tau = t/\Delta t$ corresponds to the frame index for time $t$.   As $i$ and $j$ can be identical, this definition includes autocorrelations.
When one has access to experimental correlation data $c^{exp} (t) $ it is possible to impose a constraint on the path ensemble distribution, leading to the Lagrange function
\begin{align}
\label{eq:langrangepathCorr}
\mathcal{L} &=  -\int \DD{ \XP} \PP[\XP] \ln \frac{ \PP[\XP]}{ \PP^0[\XP]}
- \nu
\left ( \int \DD \XP \PP[\XP] - 1\right )
\notag \\ &- \sum_\tau \mu_\tau \left( \int \DD\XP  \PP[\XP]
  s_i(0) s_j(t)  - c^{exp}(t)\right).
\end{align}

Following the same reasoning as for the MaxEnt approach we optimize the Lagrange function 
\begin{equation}
\elabel{eq:LagrEntpath_der}
\frac{\delta \mathcal{L} }{\delta\PP[\XP]} = -\ln \frac{ \PP[\XP]}{
  \PP^0[\XP]}  -1  - \sum_\tau \mu_\tau    s_i(x_0) s_j(x_\tau)  
 - \nu, 
\end{equation}
giving rise to the posterior
\begin{equation}
\elabel{eq:PostMaxCalCorr}
\PP^{MC}[\XP] \propto e^{- \sum_\tau \mu_\tau 
  s_i(x_0) s_j(x_\tau) }
\PP^{0}[\XP]. 
\end{equation}

As an example, suppose that we are interested in a mobility function $K_\tau[\XP]$, measuring, for example, the mean square displacement at a particular time $\tau$ with respect to time $\tau=0$. As this correlation only has to be constrained at $\tau$, the posterior is simply
\begin{equation}
\elabel{eq:PostMaxCalK}
\PP^{MC}[\XP]  \propto e^{- \mu_\tau 
  K_\tau[\XP] }\PP^{0}[\XP]. 
\end{equation}
Note that this is identical to the expression for the s-ensemble
(with $\mu =s$), which biases
path ensembles according to a time correlation function \cite{Hedges2009} and is usually presented in the context of large deviation theory. 
The s-ensemble biases all paths with a field $s$ conjugate to the function $K$.
In the MaxCal approach the Langrange multiplier $\mu$ follows from the constraint imposed. Thus, the s-ensemble might also be interpreted as the field that imposes a certain constraint. 

In any case, the posterior MaxCal distribution can be written as 
\begin{equation}
\elabel{eq:PostMaxCalgeneral}
\PP^{MC}[\XP] \propto e^{- \mu f[\XP]} \PP^{0}[\XP], 
\end{equation}
with $f[\XP]$ a function of the path $\XP$ imposing the constraint. 
Combining \eref{eq:PostMaxCalgeneral} with the path probability $\PP [\XP] = \exp(-\mathcal{S}[\XP])$, in terms of the path action $\mathcal{S}[\XP]$ 
gives
\begin{equation}
\label{eq:PMCACTIOn}
\PP^{MC}[\XP]   \propto \exp(-\mu  f[\XP]) \exp(-\mathcal{S}^0[\XP]), 
\end{equation}
which can be rewritten as 
\begin{equation}
\elabel{eq:SMC}
S^{MC}[\XP]
= \mu  f[\XP]  + \mathcal{S}^{0}[\XP], 
\end{equation}
\eref{eq:SMC} quantifies the correction of the prior action $\mathcal{S}^{0}$
by the experimental constraints.


\subsection{Thermodynamic constraints}
Since equilibrium properties are not time dependent, they can be computed as time averages over path ensembles distributions:
\begin{equation}
\elabel{eq:thermo1}
\langle s \rangle =  \frac{1}{\langle L \rangle}  \int \DD{\boldsymbol \XP}
\PP[\XP] \sum_t s(x_t),
\end{equation} 
with $\langle L \rangle$ being the average path length, and $x_t$ the coordinates at each timestep of the path. 
Constraining an equilibrium property $s^{exp}$ then
leads to a posterior distribution 
\begin{equation}
\elabel{eq:PostMaxCaltime}
\PP^{MC}[\XP]   \propto e^{- \mu 
\sum_t s(x_t)}
\PP^{0}[\XP].  
\end{equation}
An alternative way of constraining  equilibrium properties is to first
reduce the path space back to a configurational density $\rho(x)\equiv P(x)$ by
\begin{equation}
\elabel{eq:thermo3}
\rho(x) \propto \int \DD{\boldsymbol \XP} \PP[\XP] \sum_t \delta( x_t -x).
\end{equation}
The average then becomes simply 
\begin{equation}
\elabel{eq:thermo4}
\langle s\rangle = \frac{\int dx \rho(x)  s(x) } {\int dx \rho(x)  }. 
\end{equation} 
Indeed, substitution of \eref{eq:PostMaxCaltime} and \eref{eq:thermo3} in \eref{eq:thermo4} yields the same result as \eref{eq:thermo1}.

\subsection{Independence of  partial path
  distributions}

Up to now we did not specify what the path ensemble distribution refers to. In what follows we focus on systems that show two-state kinetics between two  stable states A and B.  We assume that there is a  separation between the molecular timescale and the reaction time\cite{Chandlerbook}, to guarantee that well-defined rate constants exist for  the interconversions between A and B. 
The total distribution $\PP[\XP] = \PP_A[\XP] + \PP_B[\XP]$ is the sum
of the (unnormalised) partial path
distributions $\PP_A[\XP] \equiv \PP[\XP]  h_A(x_0)  $ and $\PP_B[\XP]
\equiv \PP[\XP]  h_B(x_0)$,  consisting respectively,
of  all paths that  start in A, and paths that start in B. 
Here $h_{A,B}(x)$ are the indicator functions, which are unity when
the configuration $x$ is in state $A(B)$,  and zero otherwise.
Note that we restrict all paths to start and end in one of the stable states. 

In the next sections we will focus on applying kinetic constraints on
each partial path ensemble
separately, as they can be treated independent from each other. To show that, we apply
two dynamical constraints onto the total distribution, one for each
 partial ensemble
\begin{align}
\mathcal{L} &=  -\int \DD{ \XP} \PP[\XP] \ln \frac{ \PP[\XP]}{ \PP^0[\XP]}
- \nu
\left ( \int \DD \XP \PP[\XP] - 1\right )
\notag \\ &- \mu_A \left( \int \DD\XP  \PP[\XP]
  h_A(x_0) s_A[\XP]  - s_A^{exp}\right)  \\ &-\mu_B \left( \int \DD\XP  \PP[\XP]
  h_B(x_0) s_B[\XP]  - s_B^{exp}\right).
\end{align}
where we used the definition of the partial ensembles.
Maximisation of the caliber yields the posterior
\begin{equation}
\PP^{MC}[\XP] \propto e^{-  \mu_A   h_A(x_0) 
  s_A[\XP] -  \mu_B   h_B(x_0) 
  s_B[\XP] }
\PP^{0}[\XP],
\end{equation}
or, expressing it in partial ensembles
\begin{multline}
    \PP_A^{MC}[\XP]  + \PP_B^{MC}[\XP] \propto e^{-  \mu_A   h_A(x_0) 
  s_A[\XP] -  \mu_B   h_B(x_0) 
  s_B[\XP] } \times \\
\times ( \PP_A^{0}[\XP] +\PP_B^{0}[\XP]). 
\end{multline}
Clearly, for  paths belonging to partial ensemble A  $h_A =1$ and thus $h_B=0$
\begin{equation}
\PP_A^{MC}[\XP]  \propto e^{-  \mu_A     s_A[\XP] }
\PP_A^{0}[\XP] 
\end{equation}
while for paths from  partial ensemble B $h_A =0$ and $h_B=1$ it holds

\begin{equation}
\PP_B^{MC}[\XP]  \propto e^{-  \mu_B    s_B[\XP] }
\PP_B^{0}[\XP] 
\end{equation}
Thus, both partial ensembles can be optimised and normalised independently.
 Indeed, when imposing kinetic constraints, this is what we aim to do.

\subsection{Constraining rate constants using MaxCal}
We now turn to constraining kinetic observables, and in particular rate constants. Suppose we have unbiased simulations that we want to correct in order to match an experimental rate constant, $s^{exp} \equiv k_{AB}^{exp}$.
First, we need to look at how the rate is defined in the path space as the time derivative of the correlation function 
$C(t) = {\langle   h_A(x_0) h_B(x_L)   \rangle }/{ \langle h_A(x_0) \rangle }$
 \begin{equation}
k_{AB} = \frac{dC(t)} {dt} = \frac{\langle   h_A(x_0) \dot{h}_B(x_L)   \rangle } { \langle h_A(x_0) \rangle },
\end{equation}
where the indicator functions 
$h_{A,B}(x)$ are unity when the frame is in state A  and B, respectively. In words, this expression computes the flux through entering the state B provided that the trajectories started in A.

To link the flux correlation function to the path ensembles and the maximum caliber approach, we will adopt the formalism of Transition Interface Sampling (TIS) \cite{vanErp2003,Bolhuis2009,vanErp2012}, which in turn is based on the framework of Transition Path Sampling \cite{Dellago1997,Bolhuis2002,Dellago2009}. 
Introducing a collective variable $\lambda(x)$ that can parameterise a hypersurface, or interface, in the configuration space, TIS defines a  set of
$n + 1$ non-intersecting such interfaces, denoted by the parameters $\lambda_0 < \lambda_1 < \dots < \lambda_n$. In  this way the rate constant can be written as \cite{vanErp2003} 
\begin{equation}
\elabel{eq:rate}
 k_{AB}= \phi_{1,0}  P_A(\lambda_{B}|\lambda_{1}),
\end{equation}
where 
the first term is the effective positive flux through the first interface $\lambda_0 = \lambda_A$, and the second term is the crossing probability of interface $\lambda_B=\lambda_n$ for all trajectories shot from interface 1 that came directly from state A in their backward integration. When evaluating the rate constant using the TIS framework, the  first term is accessible through a regular molecular dynamics simulation, and the second term through performing sampling the interace path ensembles using the TIS algorithm\cite{vanErp2003}, or, as an approximation by the VIE-TPS algorithm. Of course, this term can in principle also be evaluated using a very long MD simulation,  although that  is naturally  not very efficient for rare events.
The crossing probability connected to each interface ensemble is expressed as a function of $\lambda$
\begin{eqnarray}
\label{eq:croshist4}
P_A(\lambda|\lambda_0) = \int \DD \XP
\mathcal{P}_A[\XP] \theta( \lambda_{max}[\XP] -
\lambda ),
\end{eqnarray}
where
$\mathcal{P}_A[\XP]$ is the now normalised (unbiased or reweighted) path ensemble distribution for
paths leaving A,
$\theta(x)$ is
the Heaviside step function, and $\lambda_{max}[\XP]$ returns the maximum
value of  $\lambda$ along the path. Here, we assumed that $\lambda$ is
monotonically increasing with $i$. 

Imposing the constraint $k_{AB}  = k_{AB}^{exp}$, now leads to the Lagrange function
\begin{align}
\label{eq:langrangepathRate}
\mathcal{L} &=  -\int \DD{\boldsymbol \XP} \PP_A[\XP] \ln \frac{ \PP_A[\XP]}{ \PP_A^0[\XP]} - \nu
\left ( \int \DD \XP \PP_A[\XP] - 1\right )
\notag \\ &- \mu_A \left ( \int \DD\XP  \PP_A[\XP] \theta( \lambda_{max}[\XP] -
\lambda_B )  - k_{AB}^{exp} \right)
\end{align}
where we have left out the flux $\phi_{1,0}$ from the rate constant contribution for notational reasons.
Following the same reasoning as before, 
we can optimize the Lagrange function 
giving rise to the posterior
\begin{equation}
\elabel{eq:PostMaxCal}
\PP_A^{MC}[\XP] \propto e^{\mu_A   \theta( \lambda_{max}[\XP] -
\lambda_B ) } \PP_A^{0}[\XP],
\end{equation}
and from the analog of \eref{eq:Gamma} 
\begin{equation}
    k_{AB}^0 e^{\mu_A} = k_{AB}^{exp}, 
\end{equation}
we obtain the value of the Lagrange multiplier $\mu_A = \ln (k_{AB}^{exp} / k_{AB}^0 )$.
Note that this equation can easily be extended to  the analog of \eref{eq:Gammaerror}
\begin{equation}
\elabel{eq:errorkab}
    k_{AB}^0 e^{\mu_A} = k_{AB}^{exp}  + \mu_A \sigma_k^2,
\end{equation}
where $\sigma_k$ signifies the level of confidence in the rate constant data. Just as for MaxEnt, one can turn the constraint condition into a restraint condition. 

The reweighting procedure can be interpreted as a bias on only the reactive AB paths that make it to the final interface $\lambda_B$, such that the total flux of paths is
obeying the kinetic rate constraint. 
However, this means that this reweighting will introduce a
discontinuity in the path distribution, as a path that is nearly reaching B, but is recrossing back to A is not reactive, and thus not reweighted. Even though these recrossing paths themselves might be rare,
such a  discontinuity is undesirable. 
For an illustration, see~\fref{figAP} in  Appendix~\ref{sec:appC}.

\subsection{Imposing the kinetic constraint for all  $\lambda$}
We can make progress by realising that the condition that the reweighted rate should be equal to the experimental rate can be generalised to all values of $\lambda$. 
The constraint of the experimental rate should in fact apply to all values of $\lambda$. That is:
\begin{align}
\mathcal{L} &=  -\int \DD{\boldsymbol \XP} \PP_A[\XP] \ln \frac{
  \PP_A[\XP]}{ \PP_A^0[\XP]} -\nu
\left ( \int \DD \XP \PP_A[\XP] - 1\right ) \notag
\\ &-  \sum_{i=1}^{n}  \mu_i \left ( \int \DD\XP  \PP_A[\XP] \theta( \lambda_{max}[\XP] -
\lambda_i )  P_A(\lambda_n|\lambda_i)  - k_{AB}^{exp}\right )  \notag
\end{align}
where the sum runs over the $n$ interfaces.
Following the usual minimisation of the Langrange function gives
\begin{equation}
\PP_A^{MC}[\XP]  \propto \exp \left[-
\sum_{i=1}^{n}  \mu_i   \theta( \lambda_{max}[\XP] - \lambda_i )
P_A(\lambda_n|\lambda_i)  \right]  \PP_A^0[\XP].  
\elabel{eq:sumPMC}
\end{equation}
This needs to obey $n$ constraints, for $k = 1.... n$
\begin{align}
\langle \theta( \lambda_{max}[\XP] -
\lambda_k )  P_A(\lambda_n|\lambda_k) \rangle  \equiv \qquad \qquad \qquad  & \notag \\  \equiv
\frac{ \int \DD\XP  \PP_A[\XP] \theta( \lambda_{max}[\XP]  -
\lambda_k )  P_A(\lambda_n|\lambda_k)  }{ \int \DD\XP  \PP_A[\XP]  }  &= k_{AB}^{exp}, 
\end{align}
or by substitution of $\mathcal{P}^{MC}_{A}[\XP] $
\begin{align}
\elabel{eq:constraints}
\int \DD\XP \PP_A^0[\XP]   \exp \left[ -
\sum_{i=1}^{n}  \mu_i   \theta( \lambda_{max}[\XP] - \lambda_i )
P_A(\lambda_n|\lambda_i)  \right]  \times \notag \\ \times \theta( \lambda_{max}[\XP] -
\lambda_k )   P_A(\lambda_n|\lambda_k) = k_{AB}^{exp},  \qquad
\end{align}
where we for this moment assumed that $ \int \DD\XP  \PP_A[\XP]   = 1$. 

We realise that the sum in the exponent is in fact only dependent on $\lambda_{max}[\XP]$ (and of course on
$P(\lambda_n|\lambda)$), but for a given system
$P(\lambda_n|\lambda)$ is a function of $\lambda$, so the sum can be written as 
\beq
-\sum_{i=1}^{n}  \mu_i   \theta( \lambda_{max}[\XP] - \lambda_i )
P_A(\lambda_n|\lambda_i) \equiv f(\lambda_{max}[\XP] ),
\elabel{eq:flmax}
\eeq
where the $P_A(\lambda_n|\lambda)$ dependence is implicit in the function $f$. The interpretation is that the weight of each path in the posterior path ensemble is entirely dependent on the $\lambda_{max}[\XP]$. We show that this is indeed correct in Appendix~\ref{sec:app1}.

\subsection{Reweighting paths ensembles using MaxCal}

Now the question is whether this biasing function in \eref{eq:flmax}
 leads to the correct behaviour in the reweighted paths ensemble (RPE), which is a way to reweight the interface ensembles into effectively the unbiased path ensemble \cite{Rogal2010}. 
We  focus again on the (normalized) partial path ensembles $\PP_A [\XP], \PP^0_A [\XP], \PP_B [\XP] $, $\PP^0_B [\XP]$ . The projection of the RPE for the crossing probability is then
\beq
 P^0_A(\lambda|\lambda_{0})=  \int \DD \XP  \PP^0_A [\XP] \theta( \lambda_{max}[\XP] -
\lambda ),
\eeq

and the projection for the  configurational density is
\beq
 \rho^0_A(\lambda ) \propto \int \DD \XP  \PP ^0_A[\XP] \sum_{k=0}^L \delta(\lambda( x_k) - \lambda). 
\eeq
The result for the partial path ensemble coming from B is similar.

Using the MaxCal path  reweighting for  the configurational density yields
\beq
 \rho^{MC}_A(\lambda ) \propto 
\int \DD \XP  \PP^0_A [\XP]
 e^{f(\lambda_{max}[\XP])}   \sum_{k=0}^L \delta(\lambda( x_k) -
 \lambda).
\eeq
For the crossing probability the reweighting is a bit more subtle. In Appendix \ref{sec:app1} we show that 
\beq
\elabel{eq:crossprob}
 P_A^{MC}(\lambda|\lambda_{0})=   \int^{\lambda}_{\lambda_n}
 R^0_A(\lambda|\lambda_{0}) e^{f(\lambda)}   d \lambda,
\eeq
where  $R^0_A(\lambda|\lambda_{0})$ is the 'reaching' histogram of paths that just reach $\lambda$
\beq
 R^0_A(\lambda|\lambda_{0})=  \int \DD \XP  \PP^0_A [\XP] \delta( \lambda_{max}[\XP] -
\lambda ). 
\eeq 
This is the proper reweighting of the crossing probabilities.
The crossing probability for B is done likewise.

\subsection{The MaxCal bias function $f(\lambda)$ follows from MaxEnt for the density}
Now the problem is to determine the function $f(\lambda)$. MaxCal does not give a solution to this problem, as it only concerns the final rate value, which is satisfied as long as the $f(\lambda_n)$ is set to the proper value. 
Indeed, a solution to the constraint equation will be correct for all functions $f$ under the condition the $f(\lambda_B)$ gives the correct kinetic constraint. This also can be seen by defining the function 
\beq
f(\lambda_i) \equiv  F_i = \sum_{j=1}^i  \mu_j P(\lambda_n|\lambda_j), 
\eeq
where the solution to the Langrange multipliers $\mu_j$ allow virtually all reasonably shaped functions $f(\lambda)$. 

\begin{figure*}
\includegraphics[width=1\linewidth]{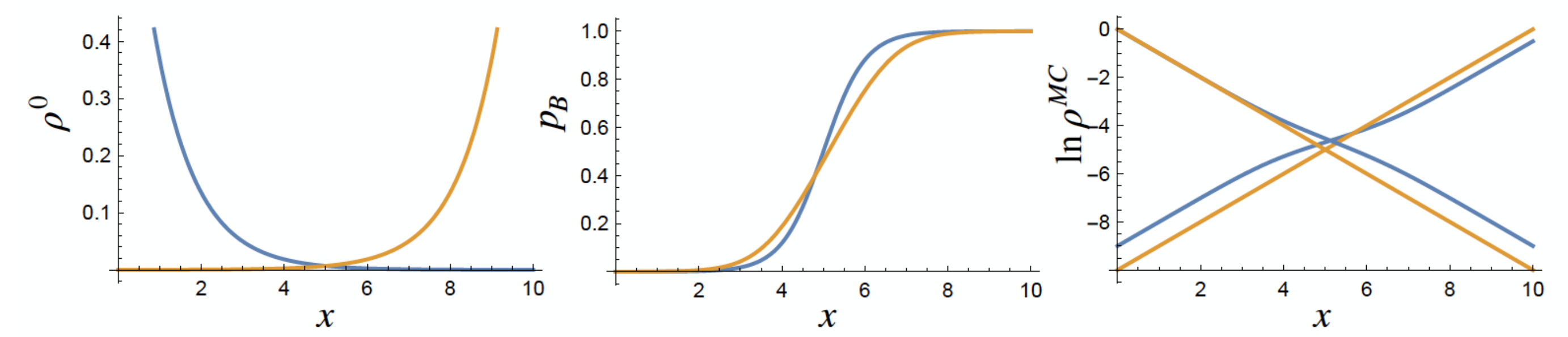}
\caption{\flabel{fig1}  Left: example initial densities $\rho_A^0
  (\lambda)$ (blue) and $\rho_B^0 (\lambda)$ (orange). Middle: initial
  committor $p_B^0 (\lambda)$ (blue). 
 Solution of the self-consistent equation
\eref{eq:implicitpb} (orange).  Right: weighted densities (blue) compared to initial densities (orange).}
\end{figure*}

Therefore, it seems that we have not made progress since $f(\lambda)$ is unknown. Here is where the configurational density comes in. For the configurational density we can also apply the regular MaxEnt approach, which yields
\begin{equation}
\elabel{eq:PostEntx}
\rho^{ME}(x) \propto e^{- \mu  g(x) } \rho^{0}(x),
\end{equation}
where $g(x)$ is a function that imposes the constraint. 
When projecting to the order parameter $\lambda$ this expression reduces to
\begin{equation}
\elabel{eq:PostEntlambda}
\rho^{ME}(\lambda) \propto e^{- \mu  g(\lambda) } \rho^{0}(\lambda),
\end{equation}
where the constraint imposed is 
\beq
\frac{\int d\lambda g(\lambda) \rho(\lambda) }{ \int d\lambda \rho(\lambda)} = g^{exp}. 
\eeq
Now what is $g^{exp}$ if we constrain the rate constants $k_{AB}$ and $k_{BA}$? The obvious candidate is the ratio $k_{AB}/k_{BA} \equiv K_{eq}$, which is
equal to  the equilibrium constant $K_{eq}=\pi_B/\pi_A$. In fact, it turns out better to consider
the equilibrium fraction $K = \pi_B/( \pi_A + \pi_B) =
K_{eq}/(1+K_{eq})$.
(Note that we used $\pi_{A,B}$ to denote the total  equilibrium population in A and B, to avoid confusion with $\rho_{A,B}(x)$).
Thus the question is which function $g(\lambda)$ would obey 
\beq
\frac{\int d\lambda g(\lambda) \rho(\lambda) }{ \int d\lambda \rho(\lambda)} = K.
\eeq
In the appendix we show that a natural solution for $g(\lambda)$ is the committor $p_B(\lambda)$, as the points that commit to B are both determining the committor, and the equilibrium fraction $K$.

The   reweighted MaxEnt densities, given in \eref{eq:PostEntlambda}, then become
\begin{eqnarray}
\elabel{eq:medensa}
\rho_A(\lambda) &=& \rho^0_A(\lambda)  e ^{\mu_A p_B(\lambda)} \\
\elabel{eq:medensb}
\rho_B(\lambda) &=& \rho^0_B(\lambda)  e ^{-\mu_B p_B(\lambda)}  e ^{\mu_A},
\end{eqnarray}
where the latter equation has a negative exponent and a shift,  and  we considered two different Lagrange multipliers, one for each direction AB and BA. 
To solve for $p_B$ we note that 
$p_B (\lambda) =  {\rho_B (\lambda)}/{(\rho_A (\lambda)+\rho_B (\lambda))}$
and substituting the ME densities gives
\beq
p_B (\lambda) =  \frac{\rho_B^0 (\lambda)}{\rho_A^0 (\lambda) e ^{-\mu_A }  e ^{(\mu_A +\mu_B) p_B(\lambda)}  +\rho_B^0 (\lambda)}.
\elabel{eq:implicitpb}
\eeq
This self-consistent equation 
can be solved numerically, given $\rho_B^0(\lambda),\rho_B^0(\lambda)$, and the values of $\mu_A$ and $\mu_B$.
These last quantities follow from the MaxCal constraint that the rate constants need to be correct.
That is 
\beq
e^{\mu_A} = \frac{k_{AB}^{exp}}{k_{AB}^0}  \hspace{2cm} e^{\mu_B}= \frac{k_{BA}^{exp}}{k_{BA}^0} 
\eeq
so that the ratio of these equations is 
\beq
e^{\mu_A- \mu_B} = K^{exp}/K^0,
\eeq
which it indeed should be.
Note that these last two equations can be extended to account for the experimental error (see \eref{eq:errorkab}). While we use MaxEnt here for clarifying purposes, we note that in principle, we can also add static constraints in the MaxCal formalism

We  illustrate this approach for a toy example density. By taking simple exponential forms for the density (see \fref{fig1}left)  we
plot the initial committor in \fref{fig1}middle. We can then apply the self-consistent solution to the committor using $\mu_A=1$ and $\mu_B=1.5$, see \fref{fig1}middle, and
reweight the densities (see \fref{fig1}right).

\subsection{Obtaining $f(\lambda)$ from $g(\lambda)$}

 What is the relation between $g(\lambda)=p_B(\lambda)$ and $f(\lambda)$? They are not identical. However, the MaxCal corrected RPE configurational density and the MaxEnt corrected configurational density should be identical, i.e.
\beq
 \rho^{MC}_A(\lambda )  = e^{- \mu  g(\lambda) }   \rho^0_A(\lambda ), 
\eeq
or 
\begin{align}
  \elabel{eq:volterra}
\int \DD \XP  \PP^0_A [\XP]
 e^{f(\lambda_{max}[\XP])}   \sum_{k=0}^L \delta(\lambda( x_k) -
 \lambda)  =  \qquad \notag \\  =   e^{- \mu  g(\lambda) }    \int \DD \XP  \PP^0_A [\XP]
  \sum_{k=0}^L \delta(\lambda( x_k) -
 \lambda). 
\end{align}
In practice, this Volterra equation of the first kind should be solved numerically. 
For instance, when the configurational density histograms $\rho^0_A(\lambda,\lambda_m )$ are computed for each interface value $\lambda_m$, the maximum value of each path in that ensemble, this equation becomes
\beq
\int_{\lambda_m=\lambda_n}^{\lambda}
 \rho^0_A(\lambda,\lambda_m )
 e^{ f(\lambda_{m})}    =   e^{- \mu  g(\lambda) }   
\int_{\lambda_m=\lambda_n }^{\lambda}
\rho^0_A(\lambda,\lambda_m ),
\eeq
or in a discrete version
\beq
\sum_{\lambda_m=\lambda_n}^{\lambda}
 \rho^0_A(\lambda,\lambda_m )
 e^{ f(\lambda_{m})}    =   e^{- \mu  g(\lambda) }   
\sum_{\lambda_m=\lambda_n }^{\lambda}
\rho^0_A(\lambda,\lambda_m ).
\eeq
This function can be solved numerically starting from the final value $\lambda_m=\lambda_n$
for which holds 
\beq
 \rho^0_A(\lambda,\lambda_n )
 e^{ f(\lambda_{n})}    =   e^{- \mu  g(\lambda_n) }   
\rho^0_A(\lambda,\lambda_n )
\elabel{eq:backsub}
\eeq
Iteration by back-substitution leads to the desired weighting function
$f(\lambda)$.
A similar equation needs to be solved for the partial ensemble of paths starting in B.

The entire procedure is summarised as an equation scheme in  \fref{fig:eqscheme}. 
\begin{figure}[t]
\includegraphics[width=1\linewidth]{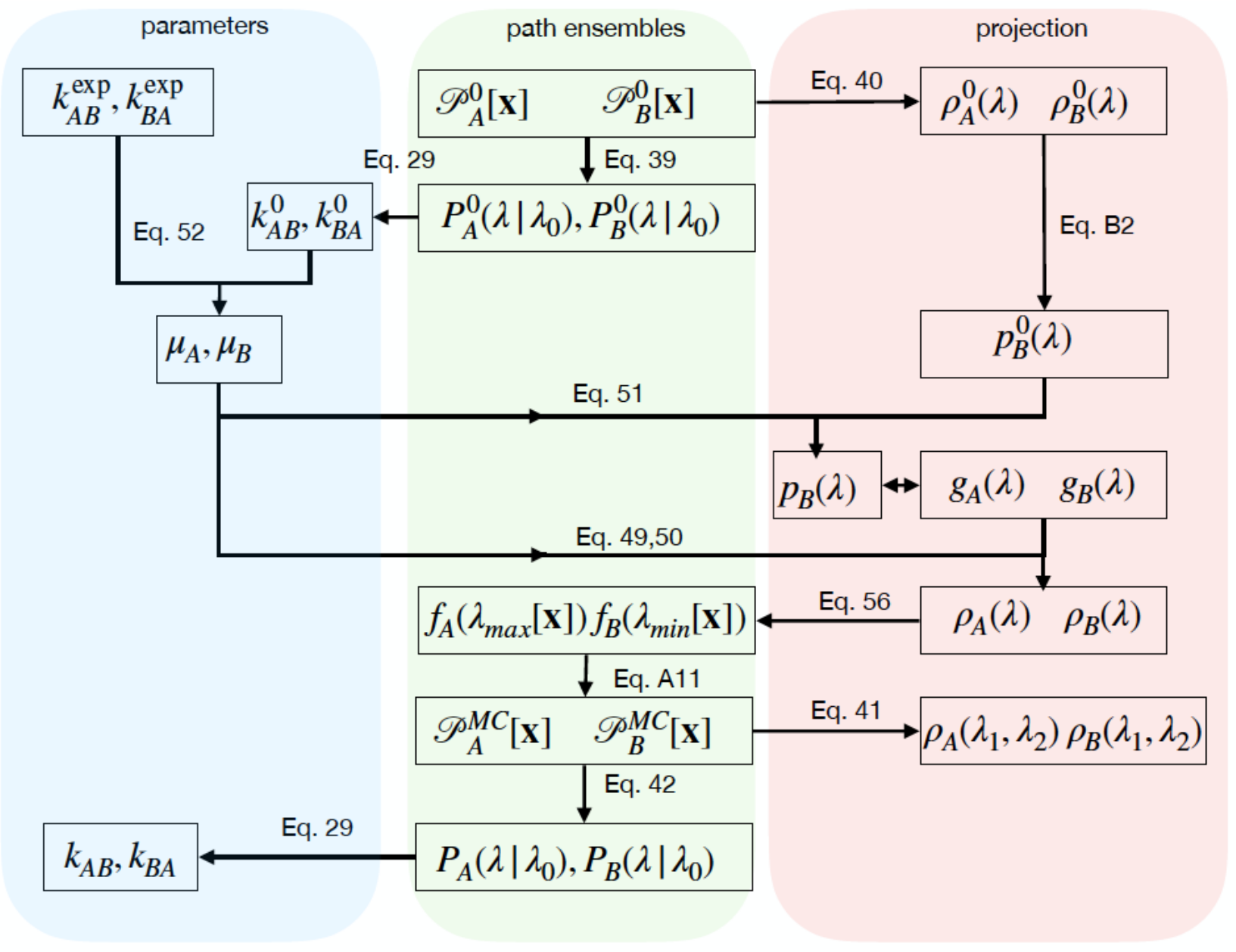}
\caption{\flabel{fig:eqscheme} Scheme relating  all important parameters, path ensembles, and projected functions  via the key equations. Starting from the top the   arrows indicate the direction and  order in which the method applies the equations. } 
\end{figure}

\subsection{Optimal path distributions by varying the CV}
\label{sec:optcv}

 The final perturbed distributions will  be dependent on the choice of the CV. In principle, it is possible to vary the CV and maximizing the entropy and caliber as function of the CV. The most optimal CV is then the one  that leads to the least perturbed path distribution. 
 
Inserting  the  optimised MaxCal
distributions  $\PP^{MC}_A[\XP] = C_A^{-1} \PP_A^0[\XP] \exp[f_A(\lambda_{max}[\XP])]$ and   $\PP^{MC}_B[\XP] =  C_B^{-1} \PP_B^0[\XP] \exp[f_B(\lambda_{min}[\XP])]$, with $C_A, C_B$ appropriate normalisation constants, into the expression  for the caliber of the distributions 
yields 
\begin{align}
 \elabel{eq:KLdivmt}
& S_A[\PP_A || \PP_A^0] =\\
&   -\frac{1}{C_A} \int \DD{ \XP}
  \PP^0_A[\XP] e^{f_A(\lambda_{max}[\XP])} (
    f_A(\lambda_{max}[\XP]) - \ln C_A)  \notag 
\end{align}
and
\begin{align}
  & S_B[\PP_B || \PP_B^0] = \\
  & -\frac{1}{C_B} \int \DD{ \XP}
    \PP^0_B[\XP]
    e^{f_B(\lambda_{min}[\XP])} (
    f_B(\lambda_{min}[\XP]) - \ln C_B) \notag
\end{align}
Using the definition of the 'reaching histograms'
$R^0_A(\lambda|\lambda_0), R^0_B(\lambda|\lambda_n) $   this becomes 
\begin{align}
S[\PP_A || \PP_A^0]&= -\frac{1}{C_A} \int d \lambda R^0_A (\lambda|\lambda_0) e^{f_A(\lambda)}
                   ( f_A(\lambda) - \ln C_A)  \notag \\ 
                                                          S[\PP_B || \PP_B^0] &=-\frac{1}{C_B} \int d \lambda R^0_B (\lambda|\lambda_n) e^{f_B(\lambda)}
(  f_B(\lambda) - \ln C_B) ,  
\end{align}
where the normalisation $C_A =  \int d \lambda R^0_A (\lambda|\lambda_0) e^{f_A(\lambda)}$, is now also expressed using the reaching histograms.
Note that we have assumed all sub-distributions $\PP_A^0, \PP_A, \PP_B, \PP_B^0$  to be normalised. However, when computing the total entropy we need to use the normalised total path distribution. 
It is possible to express the caliber for the full distributions  in terms of $S[\PP_A || \PP_A^0] $ and $S[\PP_B || \PP_B^0] $ as 
\begin{align}
  \elabel{eq:kldivalpha}
S[\PP|| \PP^0] &= \alpha S[\PP_A || \PP_A^0] + (1-\alpha) S[\PP_B || \PP_B^0]  \notag \\ &++ \alpha \ln \frac{\alpha}{\alpha_0} + (1-\alpha) \ln \frac{1-\alpha}{1-\alpha_0}
\end{align}
with $\alpha= C_A/(C_A +C_B)$, and $\alpha_0= C_A^0/(C_A^0 +C_B^0)$. The last two terms provide a kind of penalty for how much the partial ensembles differ in their respective weight. For a symmetric potential, identical sampling and a symmetric bias, $\alpha=1/2$ and  these terms vanish.

\subsection{Generalising the approach}

When deriving the $g(\lambda)$  function we use $\lambda$ as an CV. 
We can generalize the approach and look for the $g(x)$ as a function of any configuration $x$.
In analogy with \eref{eq:medensa} and \eref{eq:medensb} the reweighted MaxEnt densities are given by: 
\begin{eqnarray}
\rho_A(x) &=& \rho^0_A(x)  e ^{\mu_A p_B(x)} \notag \\
\rho_B(x) &=& \rho^0_B(x)  e ^{-\mu_B p_B(x)}  e ^{\mu_A}.
\end{eqnarray}
To solve for $p_B$ we use again the definition 
$p_B (x) =  {\rho_B (x)}/{(\rho_A (x)+\rho_B (x))}$
and substitute the MaxEnt densities,
\beq
p_B (x) =  \frac{\rho_B^0 (x)}{\rho_A^0 (x) e ^{-\mu_A }  e ^{(\mu_A +\mu_B) p_B(x)}  +\rho_B^0 (x)}.
\elabel{eq:implicitpbx}
\eeq
Again, this self-consistent equation 
needs to be solved
numerically, given $\rho_A^0(x),\rho_B^0(x)$, and the values of
$\mu_A$ and $\mu_B$.

The  $f(x)$ function then follows from identifying the MaxCal corrected RPE configurational density with the MaxEnt corrected configurational density 
\beq
 \rho^{MC}_A(x )  = e^{- \mu  g(x) }   \rho^0_A(x ) 
\eeq
or,  setting  $g(x)=p_B(x)$, 
\begin{align}
\int \DD \XP  \PP^0_A [\XP]
 e^{f(p_{B,max}[\XP])}   \sum_{k=0}^L \delta( x_k -
 x)  =  \qquad \notag \\  =   e^{- \mu  p_B(x) }    \int \DD \XP  \PP^0_A [\XP]
  \sum_{k=0}^L \delta( x_k -
 x) 
\end{align}
where $p_{B,max}[\XP]$ is the maximum value of the committor along  the
path $\XP$.
In practice this equation should be again solved numerically.

This approach is consistent with the idea that $p_B(x)$ is the most optimal reaction coordinate (RC).

\begin{figure}[b]
\includegraphics[width=1\linewidth]{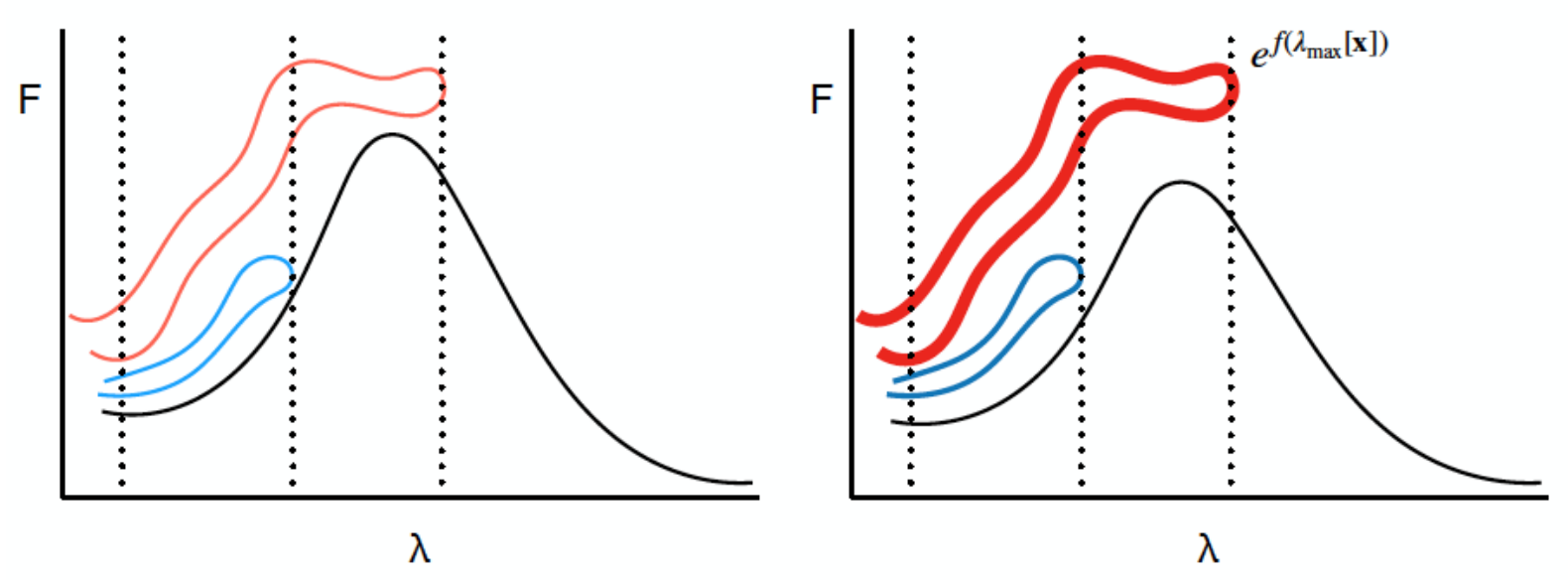}
\caption{\flabel{fig:illustration}  Illustration of how the method reweights paths. Left: the black curve depicts a free energy barrier. Several paths are shown. Blue path has a high weight, red path a low weight as it has to travel further up the barrier. Maximum $\lambda$-values are indicated by dotted vertical lines. Right: after the reweighing the red path is relatively more abundant in the ensemble, indicated by the thicker curve. The resulting free energy barrier is lowered in line with the kinetic constraints.     }
\end{figure}

\subsection{Interpretation of the method}

While the above sections give all the details of  our framework, it might be good to take a step back and summarize what is actually done.
Simply speaking, the method takes as an input the unbiased  ensembles of paths leaving state A and B, and reweights each trajectory in the ensemble according to how far it progresses along a predefined collective variable (see \fref{fig:illustration} for a equation scheme and illustration of the reweighting). This includes the paths that cross the barrier and reach the other state, so the rate constants are automatically constrained to the correct value. The more involved part of the framework is to also ensure the thermodynamic  properties are correct, in particular the equilibrium constant. This requires a specific bias function based on the committor function, which produces the least perturbed path ensemble, while still obeying the constraints. 
The interpretation of the reweighting procedure is that trajectories are artificially made more (or less) probable in the path ensembles, analogous of changing the weight of each conformation in the Boltzmann distribution, using the MaxEnt approach.  
Indeed, the reweighting would then also correspond to altering the underlying force field, leading to both different static and dynamics properties. How the force field needs to be changed to achieve this, is a different question, and might be the subject of future research.

\section{Results and Discussion}
\label{sec:results}

In this section we illustrate the approach on several toy models as well as all atom molecular dynamics simulations of association/dissociation and folding/unfolding reactions.

\begin{figure}[b]
\includegraphics[width=0.8\linewidth]{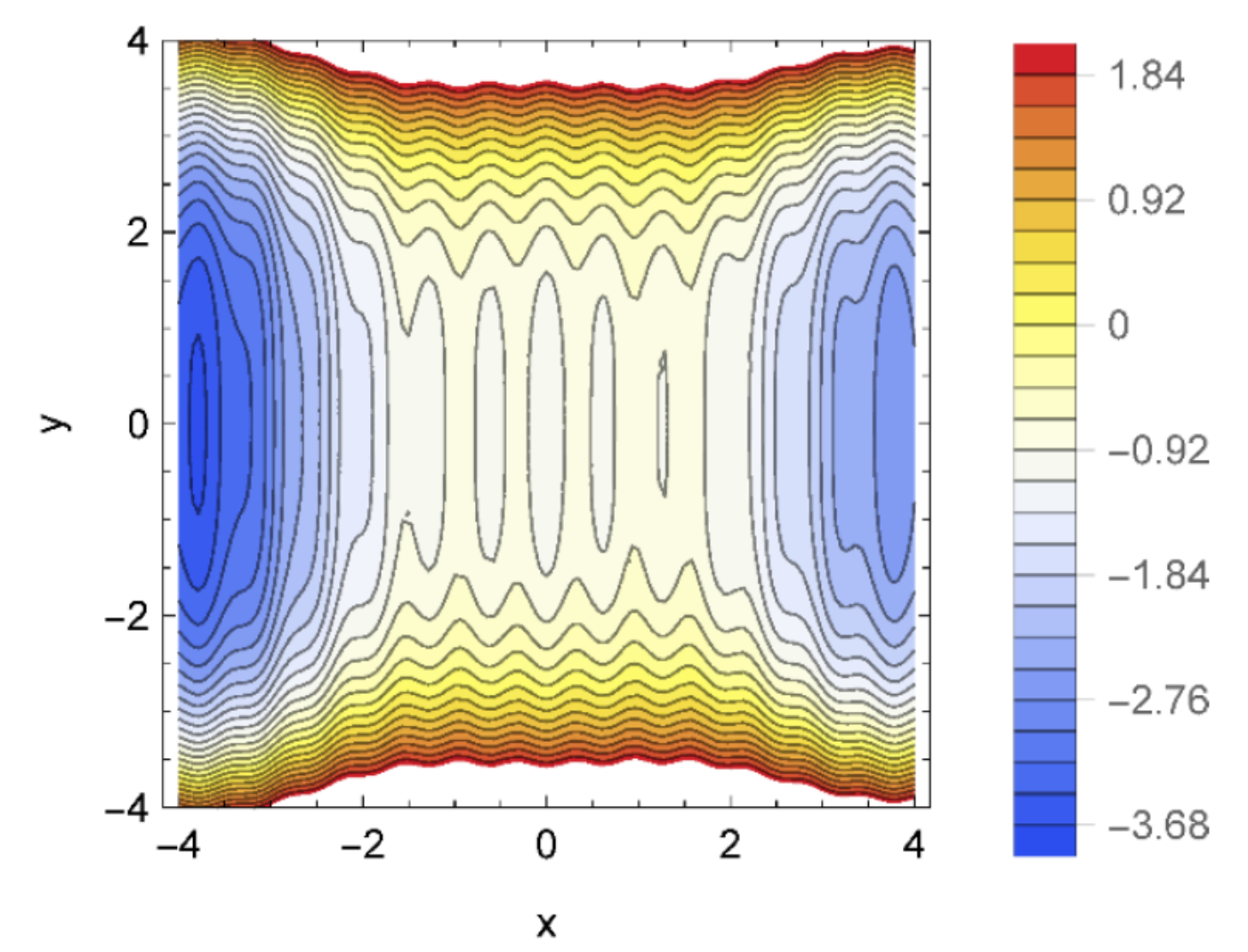}
\caption{\flabel{figpot}2D toy potential from Ref\cite{Brotzakis2019e}. Energies are in units of $k_BT$. The two states A and B are separated by an energy barrier along the x axis. Oscillations are added to show better resolution of the projections.  }
\end{figure}

\begin{figure}[t]
\includegraphics[width=1\linewidth]{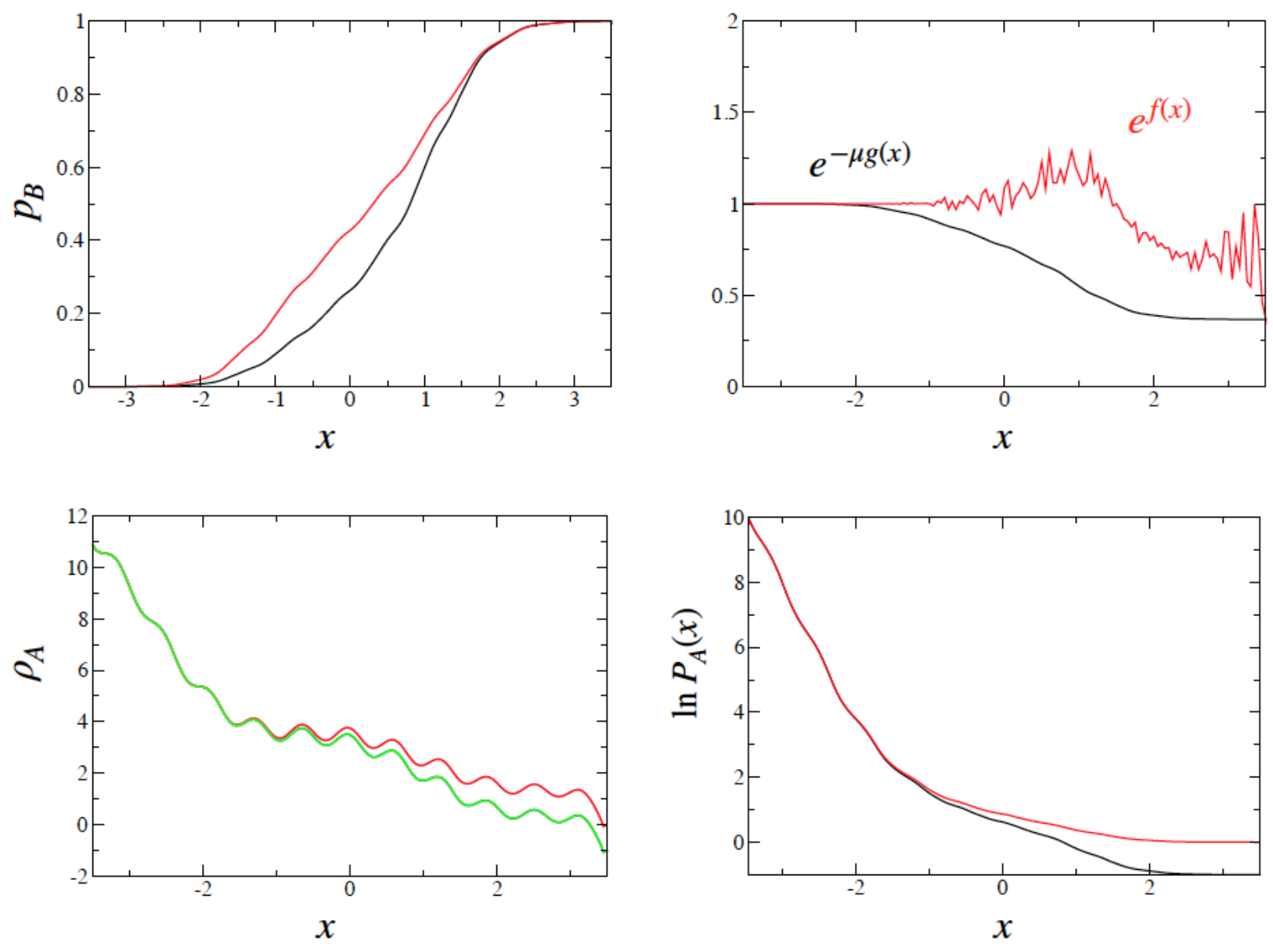}
\caption{\flabel{fig2}  Analysis of the 2D potential of~\fref{figpot}. Top left: committor $p_B^0(\lambda)$ function
  (red) and solution of the self-consistent~\eref{eq:implicitpb} (black) for
  the explicit simulation Top right: original weight function  $e^{- \mu  g(\lambda)}$ (black) and  back-iterated function  $e^{  f(\lambda)}$ (red).
Bottom left: logarithm of the configurational densities with the original in red, the reweighted with the g function (green) and the RPE-corrected with $e^{ f(\lambda)}$  in black (not visible, behind green). Bottom right: logarithm of the crossing probability with the original in red and the RPE- correct with $e^{  f(\lambda)}$ in  black.} 
\end{figure}

\begin{figure}[b]
\includegraphics[width=1\linewidth]{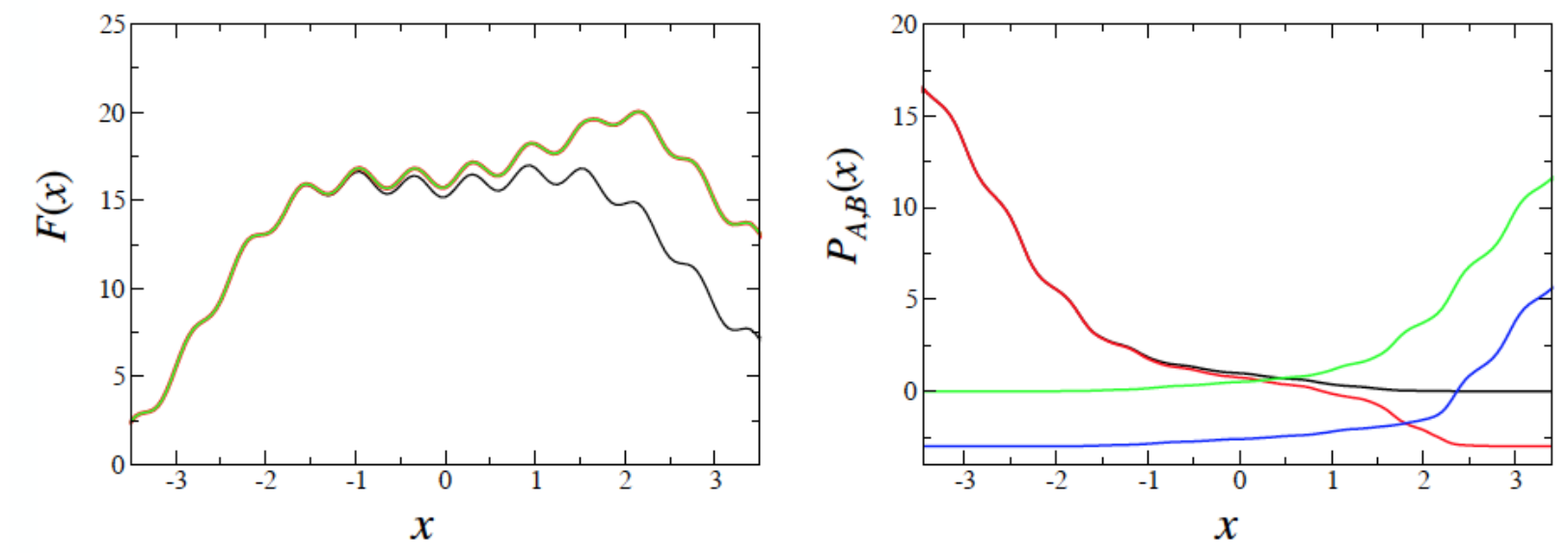}
\caption{\flabel{fig3} Analysis of the 2D potential of~\fref{figpot} by increasing the bias. Left: the free energy  for a tilt with $\mu_A=-3, \mu_B=3 $: original (black), reweighted with g function (green), RPE-corrected with $e^{ f(\lambda)}$ (red) (not visible, behind green). Right: log of the crossing probability: original in black/green, RPE-corrected with $e^{  f(\lambda)}$ in red/blue.}
\end{figure}

\subsection{\bf RPEs for toy models}
We first investigate 
a 
2D potential (\fref{figpot}) which was recently studied using the VIE-TPS method~\cite{Brotzakis2019e}. For details about the potential and the method  we refer to Ref.~\cite{Brotzakis2019e}.  Setting the (reciprocal) temperature $\beta = 1/k_BT =3$, we performed $10^7$ trials shots, where paths were generated by Metropolis Monte Carlo, on average roughly 1000 frames long. 
Applying the VIE-TPS method on this potential  gives the two partial path ensembles $\PP^0_A$ and $\PP^0_B$. We then apply our MaxCal approach,  reweighting with $\mu_A=-1$ and $\mu_B=0$, which corresponds to the lowering of the rate $k_{AB}$  by a factor $e$.
The results are shown in \fref{fig2}. The top left panel shows the
committor based on the original data  (red curve), as well as 
the self-consistent solution to the committor (black curve). 
The top right panel of \fref{fig2} gives the solution to Volterra \eref{eq:volterra} using the back substitution. 
The original weight  $e^{- \mu  g(\lambda)} = e^{- \mu  p_B (\lambda)} $ is show in black, the back iteration in red. Note that the red curve oscillates, due to numerical inaccuracies. 

Next, we show the reweighted densities $\rho_A$ in the bottom left panel of \fref{fig2}. The original density is shown in red, the reweighted with the $e^{- \mu  g(\lambda)}$ is shown in green. The RPE-corrected density  should be identical and is shown in black (Note that this is not visible as it is indeed exactly the same as the green curve).  Finally, we show the logarithm  of the  crossing probability in the bottom right panel, with the original curve in
red, and the RPE-corrected one in black. Indeed the final rate is lowered with 1, as imposed. 

For positive bias $\mu_A>0$ this treatment is also possible, but can result in some negative weights $e^{  f(\lambda)}$ for $\lambda$  just below $\lambda_n$. We  ameliorate this by putting the weights to zero for these cases, which precludes an precise solution for these cases. Still, the reweighted densities are almost correct. In any case, the values of $f(\lambda)$ do not affect the densities strongly at these values. 

In \fref{fig3} the bias is increased to $\mu_A=-3$ and $\mu_B=3$ and the crossing probabilities now shows a dramatic change. Both forward (AB) and backward (BA) curves are shown. The crossing probability curves are shifted to match the minimum values. Note that the BA curve
(blue) is thus shifted by $6 k_BT$. The free energy (left) is also shown, showing a strong shift of the transition state toward the final state.

\begin{figure}[t]
\includegraphics[width=1\linewidth]{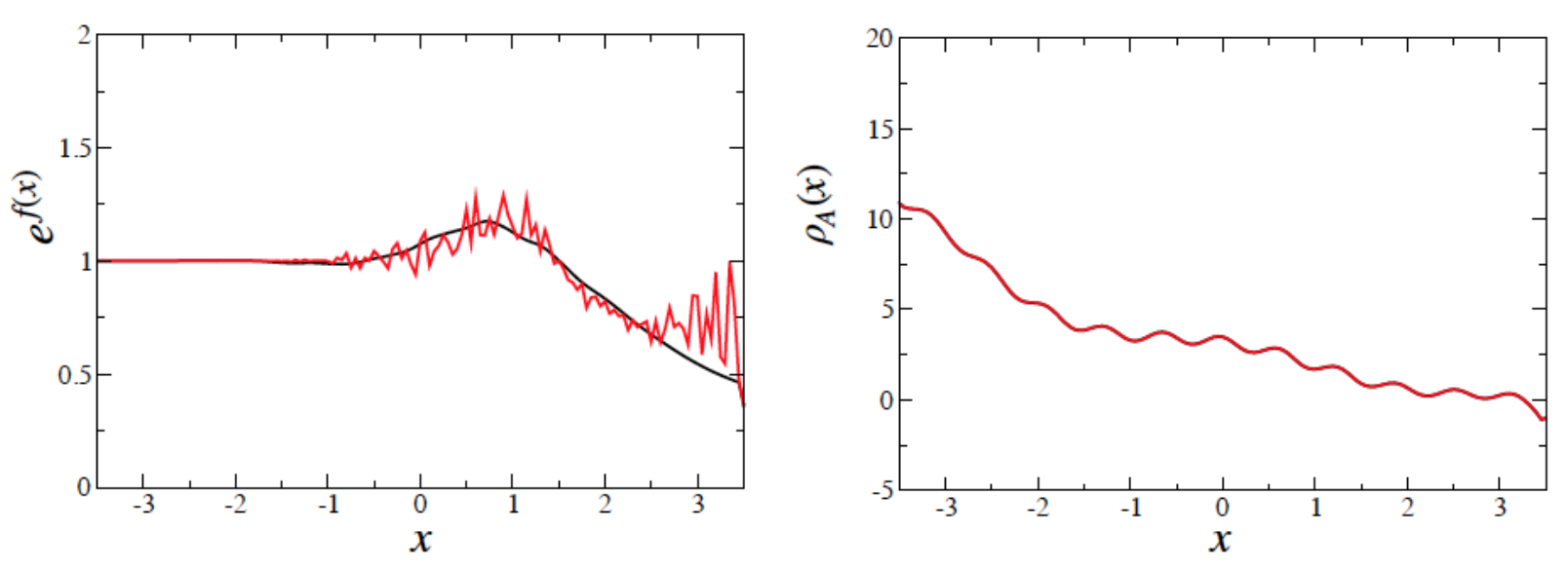}
\caption{\flabel{fig4}Analysis of the 2D potential of~\fref{figpot} by parametrizing $e^{f(\lambda)}$ with a functional form. Left:
   back-iterated function   $f(\lambda)$ (red), and the fit (black curve). Right: the corresponding densities are identical. } 
\end{figure}

The oscillations  occurring in \fref{fig2}b are related to numerical inaccuracies during the backward substitution solution for $f(\lambda)$. These oscillations indeed decrease with the amount of path ensemble data that is available. In the limit of infinite amounts of data this curves should be smooth. It should be therefore possible to parametrize $f(\lambda)$ with a functional form,
e.g. with 
$$ 
f(\lambda) = g(\lambda) + \sum_i^{n_p} a_{i,0} \exp(-a_{i,1} (x- a_{i,2})^2), 
$$
with $n_p$ the number of Gaussian functions, and optimized the $a_{i,j}$ coefficients in order to optimally accommodate \eref{eq:backsub}. The result for  $n_p=1$ is shown in \fref{fig4}. This opens up the possibility to optimize parameterisations of $f(\lambda)$ using advanced regression procedures, and even machine leaning approaches.

\subsection{Influence of the choice of CV} 

In this section we explore the influence of the choice of CV  on the optimisation. We examine two different 2D potentials which are shown in \fref{fig:2Dpots} employing standard replica exchange transition interface sampling (RETIS).
These  potentials are of the  form 
\begin{align}
v(x,y) &= 10 e^{-a  ((x+b)^2 +  (y-b)^2} \notag\\
  &-3   e^{-0.3 (x-y)^2-0.3 (x+y-8)^2}-3 e^{-0.3 (x-y)^2-0.3 (x+y+8)^2} \notag\\
& +\frac{32}{1800} \left(0.00625 (x+y)^4+10 (x-y)^2\right) . 
\end{align}
The left potential is obtained by setting $a=0.1$, and $b=2$, while the right potential  is defined by $a=0.5$, and $b=0$.

\begin{figure}[t]
  \includegraphics[width=1\linewidth]{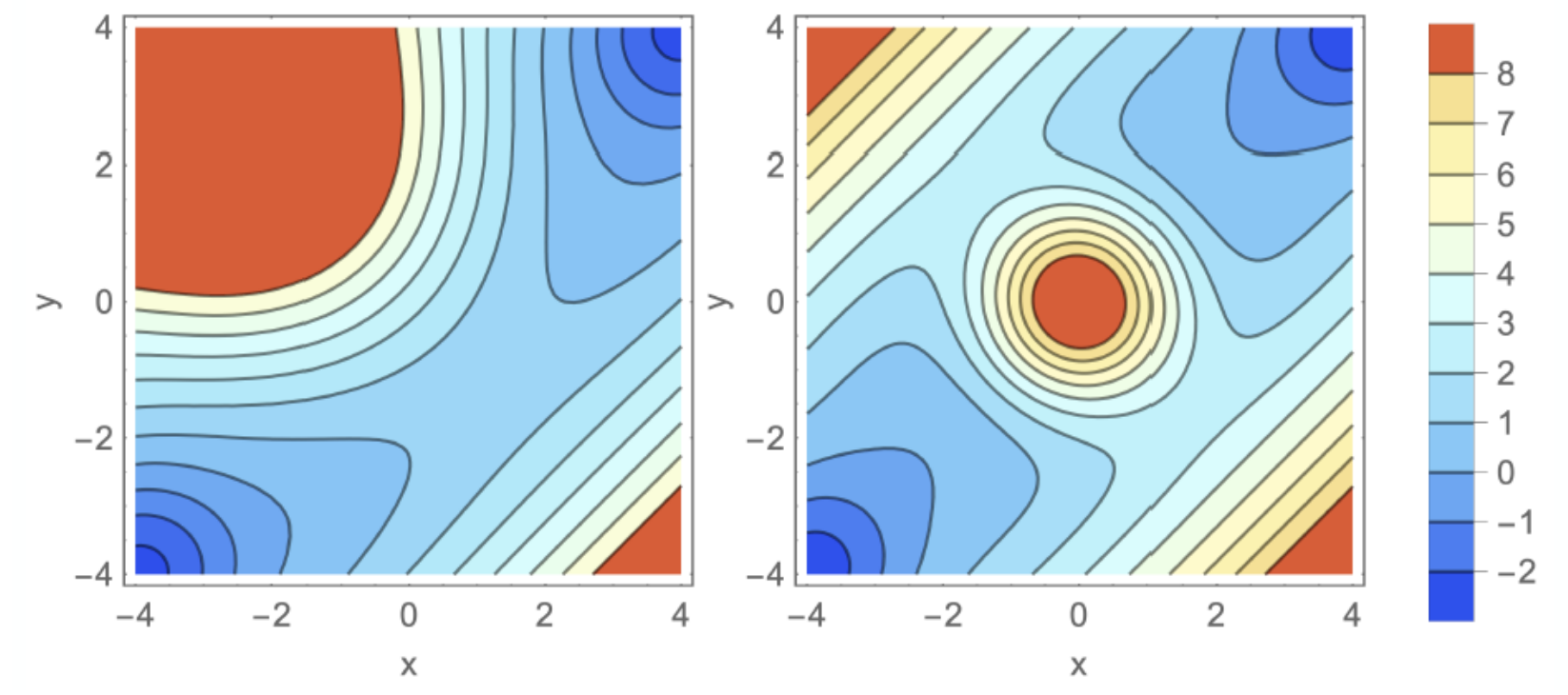}
  \caption{\flabel{fig:2Dpots} 2D toy potentials to study the influence of the order parameter.} 
\end{figure}

\begin{figure}[b!]
  \includegraphics[width=0.95\linewidth]{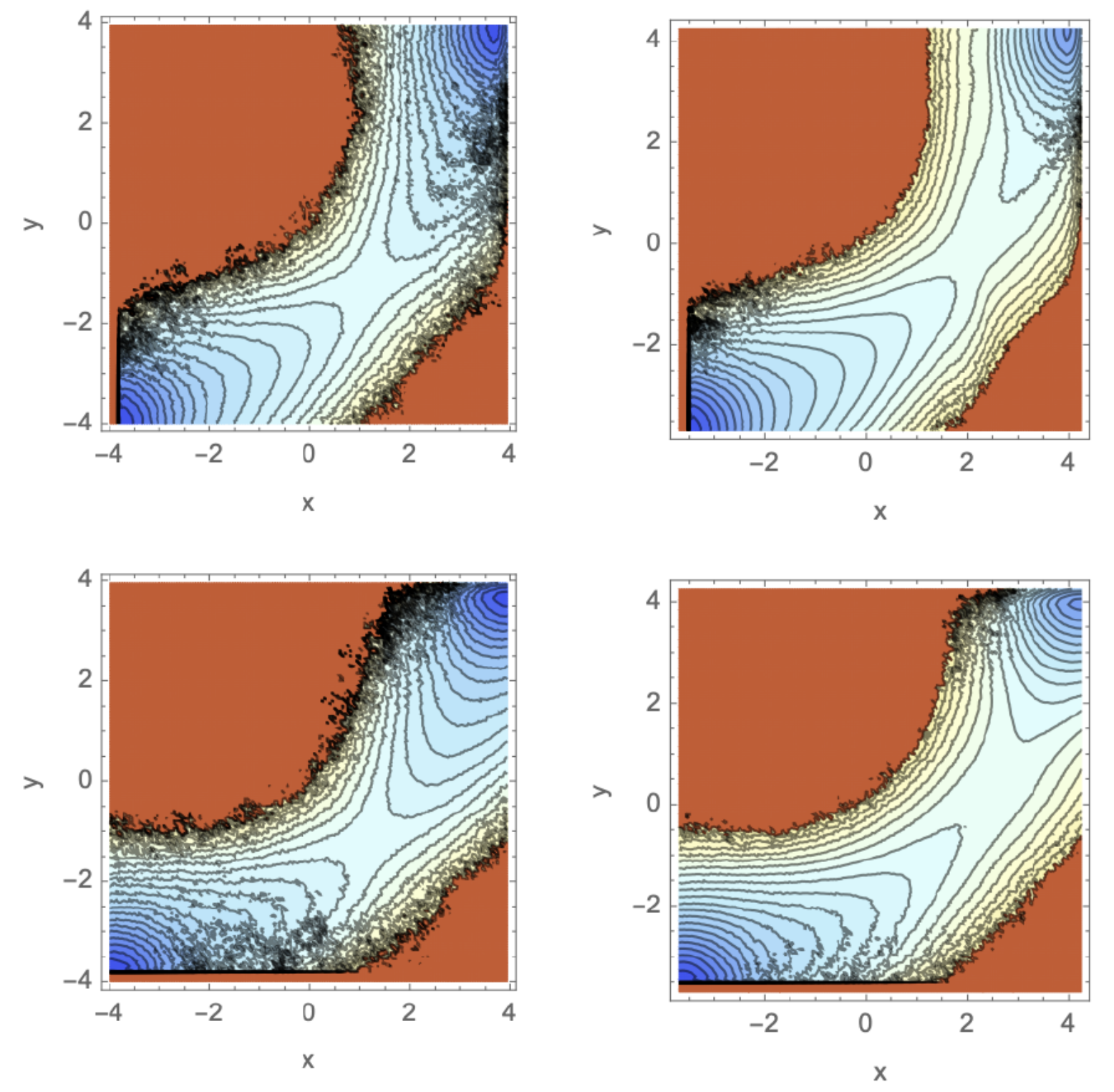}
    \caption{\flabel{fig:MCRPE} Analysis of the influence of the choice of the order parameter on the RPEs for a curved potential (\fref{fig:2Dpots}a). Free energy landscapes from the RPE for the curved potential. Left is the original sampling, with the x coordinate (top) and the y coordinate (bottom) to place interfaces. The right column shows the free energy landscape from the MaxCal posterior RPEs.}
\end{figure}
For the first potential (\fref{fig:2Dpots}a ) we run a path sampling simulation with RETIS using the $x$ coordinate
to define the interfaces. Here, we have set  29 interfaces, at respectively, $x$ values of 
$\{$-3.7,
-3.55,
-3.4,
-3.2,
-2.9,
-2.6,
-2.2,$
$-1.8,
-1.5,
-1.2,
-1.0,
-0.8,
-0.5,$
$0,
0.5,
0.8,
1.0,
1.2,
1.5,$
$1.8,
2.2,
2.6,
2.9,$
$3.2,
3.4,
3.55,
3.7$\}$. The reciprocal temperature was set $\beta = 1.5$. 
We performed $10^5$ trials shots per interface, and included replica exchanges and path reversals. Just as before we obtained the densities, which reproduced the potentials (not shown). The reciprocal temperature was set $\beta = 1.5$ 

Next, we apply the MaxCal approach with a strong kinetic constraint 
of $\mu_A= -3$ and $\mu_B=3$;  the reconstructed RPE free energy is shown in
 \fref{fig:MCRPE}.
Then we  run a path sampling simulation with RETIS using the $y$ coordinate to define the interfaces; the corresponding RPE free energy landscape is also shown in \fref{fig:MCRPE}bottom. The free energy landscape is clearly shifted,
and behaves rather independent from the choice of collective variable as order parameter for the RETIS.

\begin{figure}[t!]
  \includegraphics[width=1\linewidth]{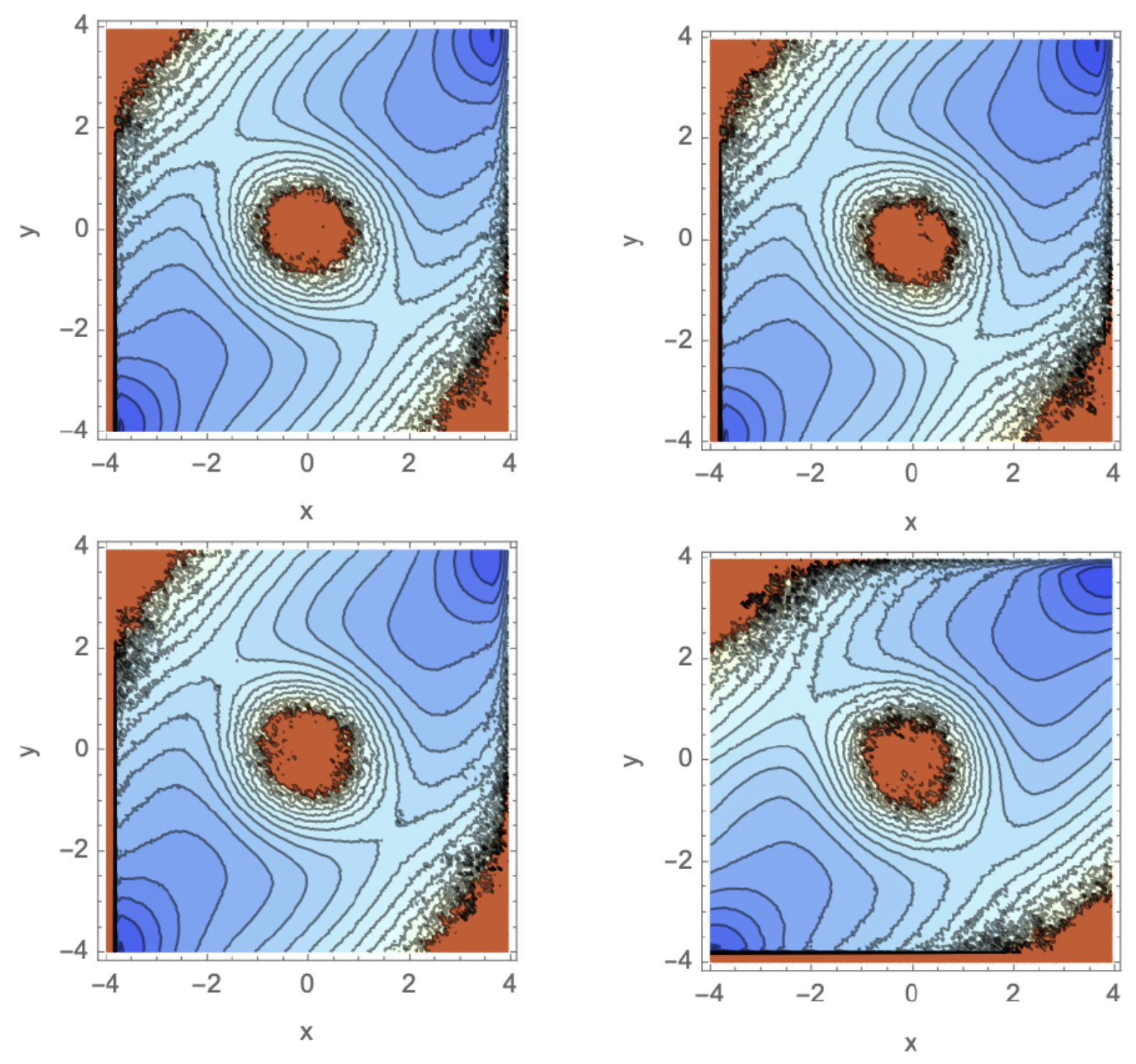}
      \caption{\flabel{fig:MCRPEtc} Analysis for the influence of choice of order parameter on the RPEs for a potential with two channels~\fref{fig:2Dpots}b. Free energy landscapes from the RPEs for the curved potential. Left is the original sampling, with the x coordinate (top) and y coordinate (bottom) to place interfaces. The right column shows the free energy landscapes form  the MaxCal posterior RPEs.}
\end{figure}

For the other  potential  with two channels we do exactly the same, albeit with a milder constraint $\mu_A= -1$ and $\mu_B=0$, and show the results in \fref{fig:MCRPEtc}b. Note that here the unbiased RPEs are the same, so it does not matter whether $x$ or $y$ is used as an order parameter to define interfaces.  
However, in the posterior FE landscapes, there are differences, depending on which CV is used as an order parameter (OP).
The upper channel is biased more when using  $y$ as an OP while the lower one is biased more when using  $x$ as an OP. This leads to a higher free energy for the upper or lower channels respectively.  This discrepancy can most likely be resolved by using the generalised approach. We leave this for a future study.

As described in section \ref{sec:optcv},  
it is possible to compute the MaxCal entropy or equivalently the KL divergence for the MaxCal optimised distributions, and identify which CV perturbs the distribution the least. Here, care needs to be taken how to add the different AB and BA sub-distributions.
Since the potential is symmetric along the diagonal, there is no  difference between the x and y CV, and both give an identical entropy or KL divergence. Therefore we compare in Table~\ref{tab:kldiv} of Appendix \ref{sec:appkldiv} the entropy of the $x$-axis with an order parameter chosen along the diagonal. 
Since the diagonal is much more aligned with the reaction coordinate, we expect the diagonal CV to give better KL divergences. And indeed, for instance for the bias $\mu_A= -1$ and $\mu_B=0$, the KL divergence for the AB path distribution is a factor of two lower for the diagonal.
When the BA ensemble is also perturbed, care needs to be taken to weight the contributions of the perturbations in the right way, as given in \eref{eq:kldivalpha}. For instance, 
for $\mu_A= -1$ and $\mu_B=1$, the BA path distribution KL divergence is at least four times smaller for the diagonal CV, as for   the x-axis CV. However, since the path distributions contribute in different proportions the total entropy for the diagonal is only twice as improved. Nevertheless, the analysis can show how the choice of the CV influences the optimisation, and how one can use this to optimize the order parameter progress variable along which the MaxCal is performed.

\begin{figure}[b!]
\includegraphics[width=1\linewidth]{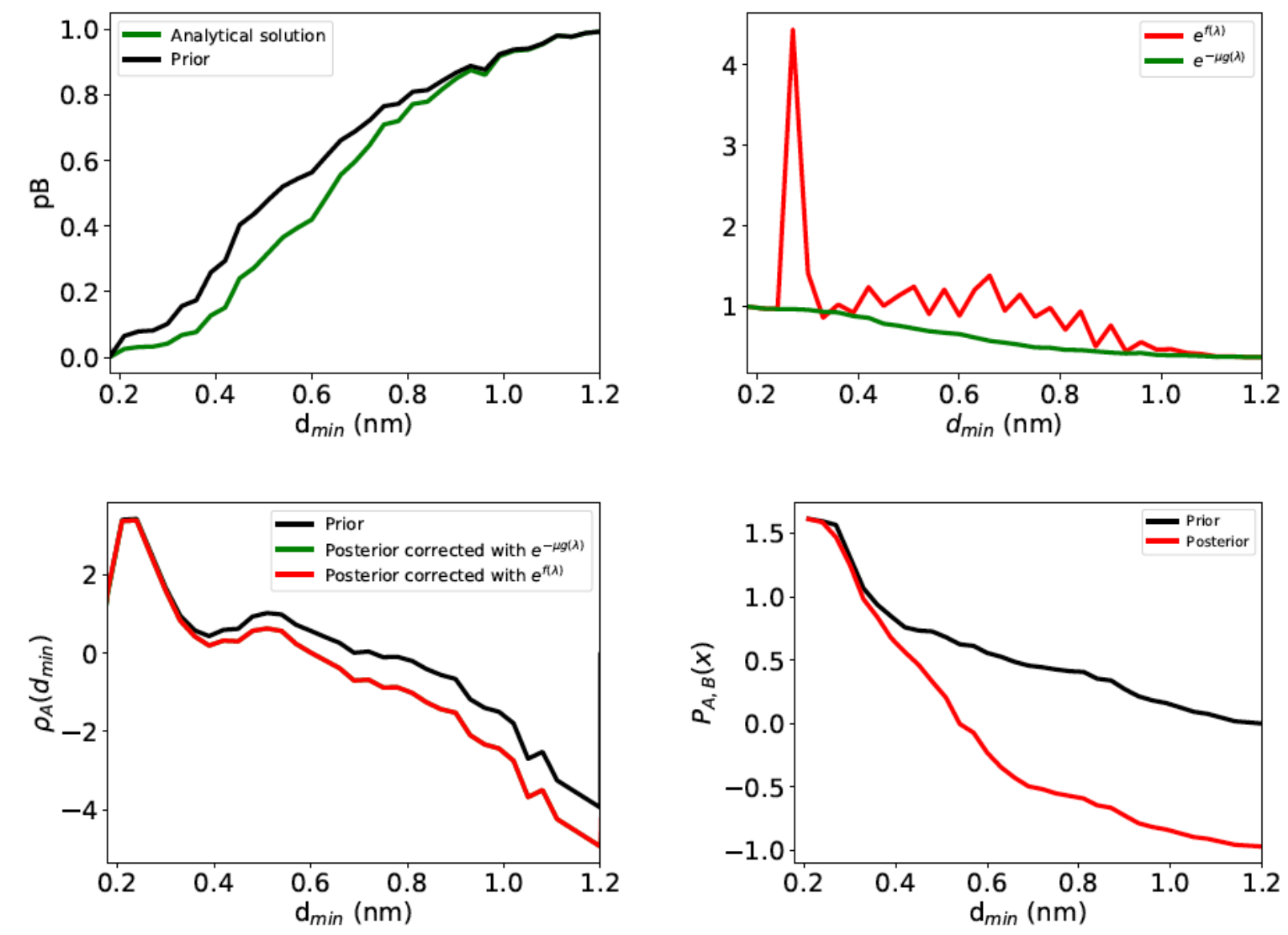}
\caption{\flabel{fig6} Dissociation of two diphenylalanine dipeptides. (a) Committor $p_B^0(\lambda)$ function (black) and solution of the self-consistent \eref{eq:implicitpb} (green) for the explicit simulation using $\mu_A$=-1. (b) Original weight function  $e^{- \mu  g(\lambda)}$ (green) and back-iterated function  $e^{  f(\lambda)}$ (red). (c) Logarithm of the configurational densities, original (black), reweighted with $g$ function (green)  (not visible, behind red), corrected with $e^{ f(\lambda)}$ (red). (d) Logarithm of the crossing probabilities, original (black), and  corrected with $e^{  f(\lambda)}$  (red).}
\end{figure}

\begin{figure}[t!]
\includegraphics[width=1.\linewidth]{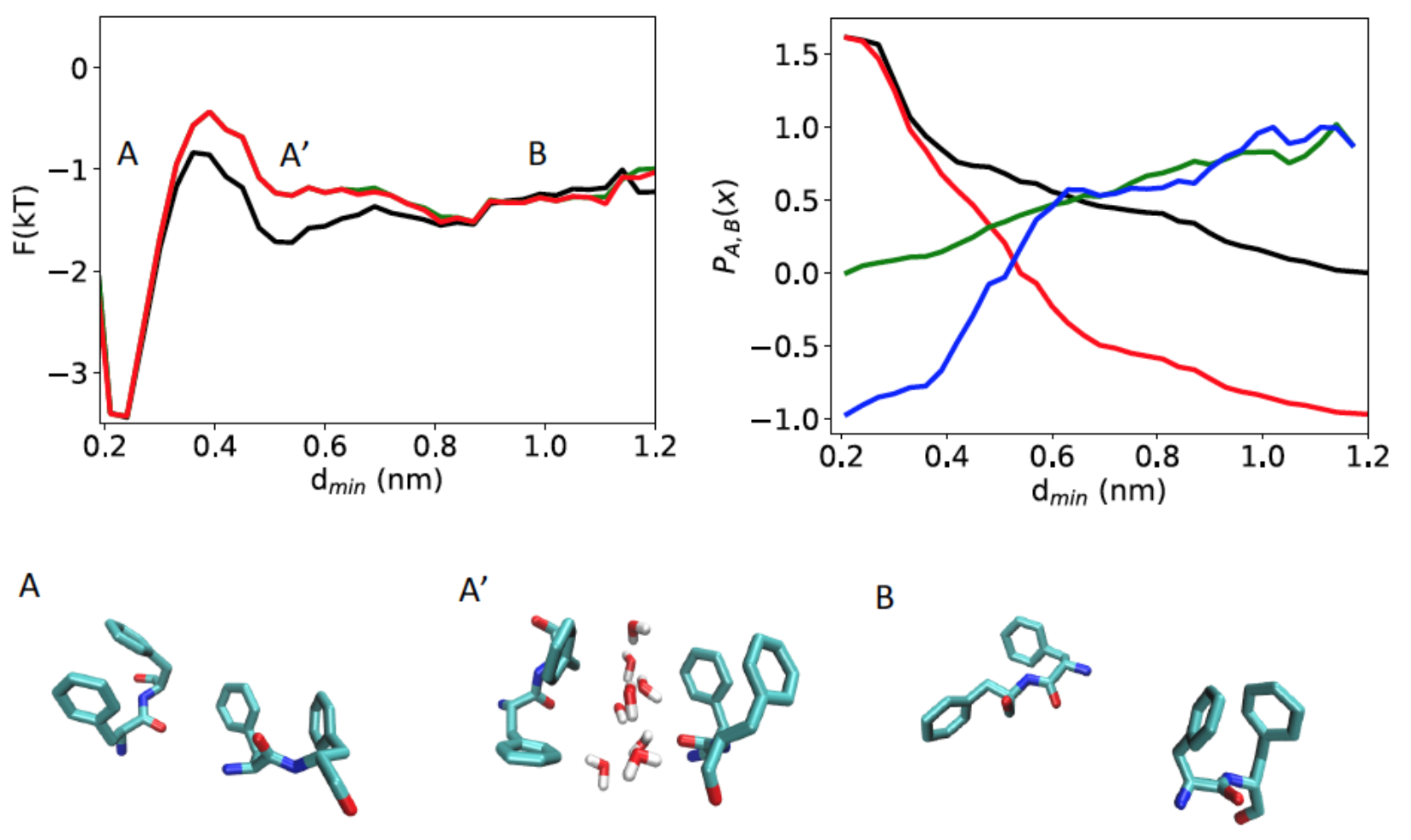}
\caption{\flabel{fig7} Association and dissociation of two diphenylalanine dipeptides. (a) Free energy  for a rate correction of
  $\mu_A=-1, \mu_B=-1 $, original (black), reweighted with $g$ function (green)  (not visible, behind red),  corrected
  with $e^{ f(\lambda)}$  (red). (b) Crossing probability histogram of original  (black/green), corrected with $e^{  f(\lambda)}$ (red/blue). c) Representative configurations of the bound (A), secondary (A') and unbound (B) states.}
\end{figure}
\subsection{\bf Association and dissociation of two diphenylalanine dipeptides}

Next, we illustrate the method to characterise the association and dissociation transition of two diphenylalanine dipeptides (FFs), as studied in Refs.~\cite{Brotzakis2016a,Brotzakis2019e}. In this system A and B refer to the bound and unbound states respectively. Here, we focus on trajectories coming from state A.
We obtain the kinetic ensemble, by using minimum distance between the peptides, $d_{min}$ as a forward model order parameter $\lambda$.
As shown in~\fref{fig6} we first reweight with $\mu_A=-1$ and $\mu_B=0$. Notably, as shown in ~\fref{fig6}a, the self-consistent solution to the committor is shown to exhibit a shift with respect to  the prior $p_B$,
and the position of the isocommittor point $p_B(\lambda)=0.5$ shifts to
larger values of the minimum distance between the peptides. The fact that we constrain $k_{AB}$ to a value of $k^0_{AB} \exp(-1)$   is reflected by the steeper posterior $p_B(\lambda)$, signifying a steeper barrier and thereby slower kinetics, as expected.

 ~\fref{fig6}b illustrates the numerical solution of the Volterra equation  \eref{eq:volterra}
 using   back substitution. The original weight  $e^{- \mu g(\lambda)} = e^{- \mu  p_B (\lambda)} $ is show in green, the back iterated solution $e^{- \mu f(\lambda)}$ in red. As for the toy models described above, the weighting function is nonlinear (red curve) and  oscillates due to numerical errors. \fref{fig6}c illustrates the original and reweighted densities $\rho_A$. The reweighted densities with the  $e^{- \mu  g(\lambda)}$ are equal to the RPE weighted ones, as expected by construction. Note the smaller probability density at values of minimum distance larger than 0.4 nm. This is exactly the effect of a smaller transition rate constant which steepens the free energy barriers. Finally, \fref{fig6}d shows the effect of the kinetic constraint to the crossing probabilities. The original crossing probability  is shown in 
red, and the RPE-corrected in black. Indeed, the crossing probability value at interface B is lowered by a factor of $1/e$, as imposed. 
In \fref{fig7} we restrain both the forward and the backward rate by $\mu_A=-1$ and $\mu_B=-1$
and the log crossing probabilities now change on both sides by a factor (-1). The reweighted densities with the  $e^{- \mu  g(\lambda)}$ are equal to the RPE-weighted ones, as expected by construction. Interestingly, making the rate exp(-1) times slower increases both the dissociation barrier ($d\approx$0.4 nm) as well as the association barrier ($d\approx$0.5 nm). The latter is done by disfavoring the stability of a secondary state $A'$ at d$\approx$0.5 nm, where one water-hydration layer is mediating peptide contacts. This water-mediated peptide-contact state $A'$ now becomes a transition state region configuration, also found in a previous study in the context of protein-protein association/dissociation~\cite{Brotzakis2019}.

\begin{figure}[b!]
\includegraphics[width=1\linewidth]{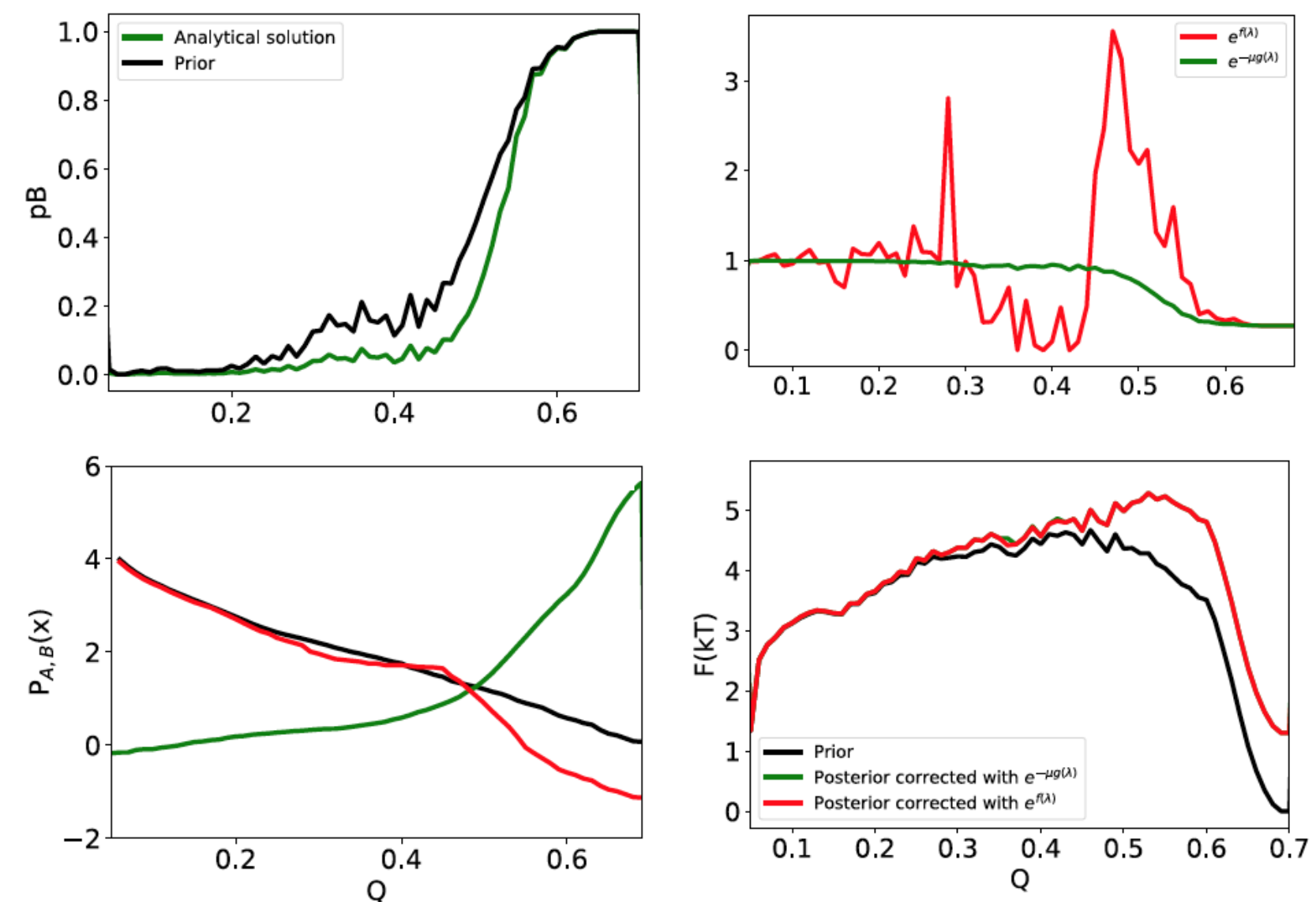}
\caption{\flabel{fig8} Simulations of folding and unfolding of chignolin. (a) Committor $p_B^0(\lambda)$ function
  (black) and solution of the self-consistent \eref{eq:implicitpb} (green) for
  the explicit simulation using $\mu_A$=-1.3. (b) Original weight function  $e^{- \mu  g(\lambda)}$ (green) and back-iterated function  $e^{  f(\lambda)}$ (red). (c) Logarithm of the crossing probability histogram of the original (black) and RPE-corrected with $e^{  f(\lambda)}$ (red). (d) Free energies as a function of the fraction of native contacts $Q$, original (black), reweighted with g function (green), RPE-corrected with $e^{ f(\lambda)}$  (red) (green not visible, behind red).}
\end{figure}

\subsection{\bf Folding and unfolding of chignolin}
The kinetics of fast folding $\beta$-hairpins have been studied by temperature-jump spectroscopy~\cite{Snow2004}, reporting
$\mu$s timescale folding. Chignolin is a
two-state, $\beta$-hairpin, mini-protein that folds in the $\mu$s timescale~\cite{Lindorff-Larsen2011}. Despite its simple fold (PDB cln025),
molecular simulation  fails to capture the experimentally-determined melting temperature of
341 K~\cite{Honda2008,Lindorff-Larsen2011,Robustelli2018}. Here we perform our kinetic analysis on  an equilibrium molecular dynamics trajectory of chignolin at 341 K~\cite{Lindorff-Larsen2011}. While at this
temperature experiments suggest that the folding and unfolding rates should be the same, simulations report a folding rate of
$k_f=1.667 \mu s^{-1}$ and an unfolding rate of  $k_u=0.455 \mu s^{-1}$ respectively. 
The corresponding enhanced stability of the folded state is likely to arise from inaccuracies in the forcefield used in the molecular dynamics simulations.
In the
absence of an experimental folding rate for chignolin, but in light of: (a) knowledge that the barrier heights should not exceed $4.5 k_B T$ ~\cite{Lindorff-Larsen2011}, and (b) that folding and unfolding rates should be the same at the melting temperature, we perform our kinetic analysis by constraining only the folding rate $k_f^{exp}=0.455 \mu s^{-1}$. This leads to posterior kinetic ensemble of (un)folding pathways exhibiting new kinetics and thermodynamics, as shown in~\fref{fig8}. 
We use the fraction of native contacts $Q$ as  
the collective variable for the order parameter $\lambda$.  In the remainder of this section states A and B refer to the unfolded ($Q<$0.05) and folded state ($Q>$0.7), respectively.

\begin{figure*}[t!]
\includegraphics[width=0.8\linewidth]{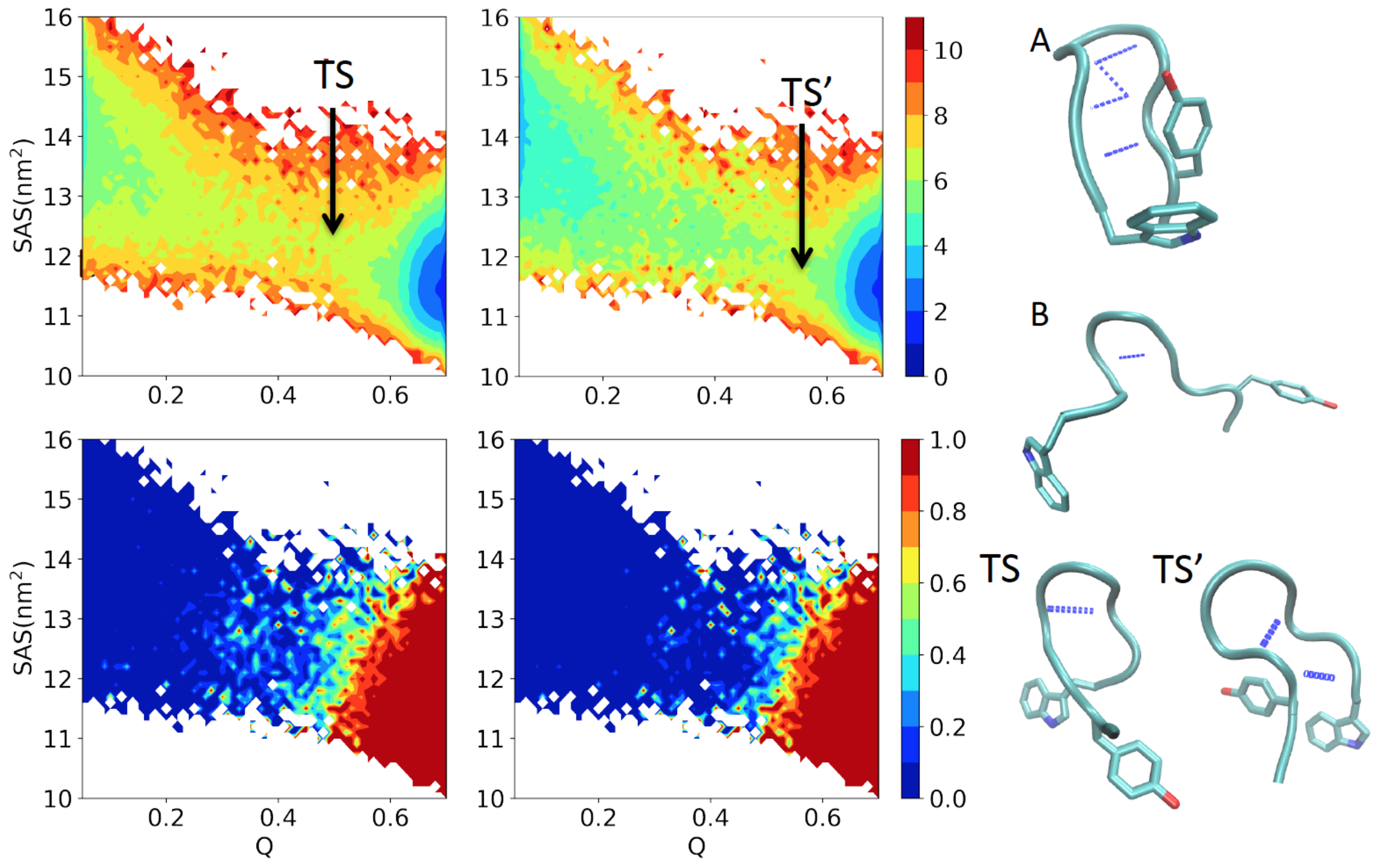}
\caption{\flabel{fig9} Free energy surface as a function of the fraction of native contacts, Q, and the solvent accessible surface, SAS, of the prior (a) and the posterior (b). In the bottom we highlight the respective committor landscapes of the prior (c) and the posterior (d). The right structure panel refers to the folded (A), unfolded(B), prior transition state (TS) and posterior transition state (TS').}
\end{figure*}

The posterior committor distribution becomes steeper and gets shifted to higher Q values (see~\fref{fig8}a). In particular the isocommittor surfaces $p_B=0.5$ shifts by 10$\%$, from $Q=0.5$ to $Q=0.55$. This is in agreement with an effect of lowering the temperature to the increase of the nativeness of the transition state~\cite{Karanicolas2002}, as well as the knowledge of native-like transition states in protein zippers~\cite{Snow2004}. 
~\fref{fig8}b illustrates the solution to the Volterra equation \eref{eq:volterra} by back substitution. The original weight  $e^{- \mu g(\lambda)} = e^{- \mu  p_B (\lambda)} $ is show in green, the
back iteration gives the MaxCal bias on the path weights $e^{  f(\lambda)}$. Applying this bias to reweight the path ensemble
results in  posterior crossing probabilities (see~\fref{fig8}c), where the folding conditional probability becomes steeper, indicating slower folding kinetics. At the same time the final shift in the folding crossing probabilities is exp(-1.3), giving indeed rise to the imposed  folding rate of $k_f^{exp}=0.455 \mu s^{-1}$. Finally,~\fref{fig8}d illustrates the effect of the kinetic constraint on the free energy. The free energy difference between folded and unfolded state becomes zero, as expected by the constraint, and amends the current force field's inaccuracy in predicting the simulated temperature (341 K) as the melting temperature. Moreover the free energy barrier becomes more asymmetric, shifting towards a more native-like conformations, as indicated also in~\fref{fig8}a.

In~\fref{fig9} we assess how the kinetic correction alters the free energy and committor landscapes as a function of fraction of native contacts and solvent accessible surface. The kinetic constraint increases the population of the misfolded configurations (0.05$<Q<$0.4) state (see~\fref{fig9}b), as well widens the distribution of solvent accessible area of the protein.  Moreover ~\fref{fig9}c,d show that the kinetic constraint shifts the transition state, i.e. the 0.5 isocommittor surface, to a higher fraction of native contacts $Q \approx  0.55$ and a slightly lower SASA of 12.2 $nm^2$, indicating a more packed structure. As illustrated in~\fref{fig9}, the prior transition state configuration TS  forms one backbone hydrogen bond and has the residues Y2 and W9,  crucial for hydrophobic collapse,  facing away from each other. On the contrary, the posterior transition state configuration TS$'$ is more native-like and shows to form more native backbone hydrogen bonds while forming contacts at the key hydrophobic collapse residues Y2 and W9.

\section{Conclusions}
\label{sec:conclusions}

Nearly 60 years after the first X-ray determination of the structure of a globular protein, molecular simulations can accurately address structural ensembles and their corresponding thermodynamics~\cite{Bonomi2017}. Yet, while the functions of proteins depend often on the transition rates between their different states, there is still a need for developing accurate methods for characterizing them.  

To address this challenge, in this work we have developed a framework to determine kinetic ensembles from experimental information. This framework combines maximum caliber and maximum entropy concepts in order to match experimentally-determined kinetic rate constants with molecular dynamics simulations. The matching is done by biasing the paths in the unbiased reweighted path ensemble based on how far they are progressing along a chosen collective variable. In this reweighting both the rate constant as well as the equilibrium free energy are constrained. 
In doing so we are able to ameliorate dynamical profiles such as conditional probabilities, committor functions and transition states, as well as the long time kinetics and the equilibrium thermodynamics. In addition, the method can impose restraints rather than constraints, by accounting for the uncertainty in the data.

To illustrate possible applications of this method we showed that matching the rate constants of protein folding of chignolin to a simulated structural ensemble yields accurate melting temperature and a more native-like transition state ensemble.


We anticipate that this method will extend the applicability of atomistic and coarse-grained molecular simulations as a kinetic tool in structural biology, e.g. for accurate mechanistic and reaction coordinate analysis.
Furthermore, the approach can be extended to amend imperfections in current atomistic force fields to reproduce the kinetics and thermodynamic observables. Such a possible method would require computing the derivative of the kinetic rate constant in path ensembles. We leave this for future research. 

We finally note that in principle the method is general and could be applied to a wide range of  problems amenable to molecular simulations.

\begin{acknowledgments}
We thank Tristan Bereau, Carlo Camiloni,  Christoph Dellago and Pieter Rein ten Wolde for carefully commenting on the manuscript. 
Z.F.B. would like to acknowledge the Federation of European Biochemical Societies (FEBS) for financial support (LTF).

\end{acknowledgments}

\appendix

\section{Constraint function} 
\label{sec:app1}
To solve the constraint equation 
\eref{eq:constraints}  
we make use of the fact that the integral over the paths
can be split into the intervals between the interfaces. For each path
with $\lambda_{max} $ between $i$ and $i+1$, the situation is then
identical. For each interval the situation is different. 
For instance, for $k=n$ it holds that only paths beyond $\lambda_n$
will contribute. 
\beq
D_n  e^{F_n}  P_A(\lambda_n|\lambda_n) = k_{AB}^{exp},
\eeq
where $D_n$ is the path fraction for paths beyond $\lambda_n$, so
$D_n = P_A^0 (\lambda_n)$, and   we have defined the sum $F_n \equiv
\sum_{i=1}^n \mu_i    P_A(\lambda_n|\lambda_i)$.   Since  $P_A(\lambda_n|\lambda_n) =1$
by definition this changes into
\beq
  e^{F_n  } = k_{AB}^{exp} /D_n =  k_{AB}^{exp} / P_A^0
  (\lambda_n|\lambda_1) = \frac{P_A(\lambda_n|\lambda_{0}) } {P_A^0 (\lambda_n|\lambda_0) } 
\equiv  e^{\mu  f(\lambda_n)}
\eeq
Indeed this is the correct reweighting that is needed.
For the case of $k=n-1$
\beq
\left(D_{n-1}  e^{F_{n-1} }   + D_n  e^{F_n} \right) P_A(\lambda_n|\lambda_{n-1}) = k_{AB}^{exp} 
\eeq
Now $D_{n-1}  =  P_A^0 (\lambda_{n-1}|\lambda_0) -P_A^0 (\lambda_n|\lambda_0) $, and
\beq
P_A(\lambda_n|\lambda_{n-1})  = P_A(\lambda_{n}|\lambda_0)
/P_A(\lambda_{n-1}|\lambda_{0})
\eeq
This will give
\beq
D_{n-1}  e^{F_{n-1}   }   + D_n
  e^{F_n }   = k_{AB}^{exp}
\frac{P_A(\lambda_{n-1}|\lambda_{0}) } { P_A(\lambda_n|\lambda_{0}) }
\eeq
or 
\beq
D_{n-1}  e^{F_{n-1}    }   + 
 k_{AB}^{exp}   = k_{AB}^{exp}
\frac{P_A(\lambda_{n-1}|\lambda_{0}) } { P_A(\lambda_n|\lambda_{0}) },
\eeq
which is 
\beq
D_{n-1}  e^{F_{n-1}  }   
 = \frac{k_{AB}^{exp} }{P_A(\lambda_n|\lambda_{0}) }  ( P_A(\lambda_{n-1}|\lambda_{0})  - P_A(\lambda_n|\lambda_{0}) ),
\eeq
or, since $k_{AB}^{exp}  = P_A(\lambda_n|\lambda_{0}) $ by definition, the weight for paths in
the $k=n-1$ slot becomes
\beq
e^{F_{n-1}   } 
 = \frac{
   P_A(\lambda_{n-1}|\lambda_{0})  - P_A(\lambda_n|\lambda_{0}) } {P_A^0
   (\lambda_{n-1}|\lambda_0) -P_A^0 (\lambda_n|\lambda_0) }. 
\eeq
This gives a regular pattern, and for $k=i$ it holds that for paths in this slot the weight is as follows
\beq
\elabel{eq:final}
e^{F{i} }  
 = \frac{
   P_A(\lambda_{i}|\lambda_{0})  - P_A(\lambda_{i+1}|\lambda_{0}) } {P_A^0
   (\lambda_{i}|\lambda_0) -P_A^0 (\lambda_{i+1}|\lambda_0) } 
\eeq

The next question is what the function $F_i$ is. This function can be
expressed as a function of $\lambda$ by noticing that all the paths
fall in a slot $i$ and $i+1$, which are determined by $\lambda_{max}
[\XP]$. This means that for paths in this slot
\beq
F_i = f(\lambda_{max} [\XP])
\eeq
This would mean that each path should reweighted using 
\beq
\elabel{eq:reweight}
\PP_A^{MC}[\XP] \propto \exp(  f( \lambda_{max}[\XP] )) \PP_A^{0}[\XP]
\eeq
where the function $f(\lambda)$ should be the biasing
function.

Now we should still check whether $ \int \DD\XP  \PP_A[\XP]   = 1$. To
do so we expand the integral in intervals, as before, to get
\begin{align}
\int \DD\XP  \PP_A[\XP]   &= \sum_i^n  D_i  \exp \left[
\sum_{i=1}^{n}  \mu_j   \theta( \lambda_{max}[\XP] - \lambda_j )
P_A(\lambda_n|\lambda_j)  \right]   \notag \\ &= \sum_i^n
P_A(\lambda_{i}|\lambda_{0})  - P_A(\lambda_{i+1}|\lambda_{0})  = 1
\end{align}
where we used the fact that $P_A(\lambda_{0}|\lambda_{0})=1$. So indeed
the normalisation is guaranteed.

At first sight it seems that, when the path histograms, e.g. the
crossing probabilities $P_A(\lambda|\lambda_0)$, are reweighted with a
function $f(\lambda)$,  we can simply replace 
\begin{align}
 P^{MC}_A(\lambda|\lambda_{0}) &= C \int \DD \XP  \PP^0_A [\XP] \theta( \lambda_{max}[\XP] -
\lambda ) e^{   f(\lambda_{max}[\XP] ) } 
\end{align}
 by $ P^0_A(\lambda|\lambda_{0})  e^{ f(\lambda_{max}[\XP] ) } $.  
But this would be wrong.
The $\theta$-function in the integral means we have to sum over ensemble
of paths for a certain $\lambda_{max}[\XP]$. 
To see this we should look at the  'reaching'  histogram :
\beq
 R^0_A(\lambda|\lambda_{0})= C \int \DD \XP  \PP^0_A [\XP] \delta( \lambda_{max}[\XP] -
\lambda ), 
\eeq
where $C$ is a normalisation constant.
This histogram counts the paths that 'just reached' $\lambda$.  The
crossing probability can be simply obtained from this by integration:
\beq
 P^0_A(\lambda|\lambda_{0})= C  \int^{\lambda}_{\lambda_n}
 R^0_A(\lambda|\lambda_{0}) d \lambda.
\eeq
This can be easily seen by realising   $\theta( \lambda_{max}[\XP] -
\lambda ) =\int^{\lambda}_{\lambda_n}\ \delta( \lambda_{max}[\XP]
-\lambda ) d \lambda$. 
For $\lambda=\lambda_n$, all paths that cross $\lambda_n$ should be
included. Effectively this could be achieved by taking $\lambda_n
\rightarrow \infty$

 When the paths are reweighed with $e^{f(\lambda_{max}[\XP])}$ 
this should be done on $R^0_A \lambda|\lambda_{0})$, and not directly
on  $P^0_A(\lambda|\lambda_{0})$. Thus
\begin{align}
R^{MC}_A(\lambda|\lambda_{0}) &= C \int \DD \XP  \PP_A [\XP]
  \delta( \lambda_{max}[\XP] -
\lambda )   \notag \\ &= C \int \DD \XP  \PP^0_A [\XP]
e^{f(\lambda_{max}[\XP])}     \delta( \lambda_{max}[\XP] -
\lambda ) \notag \\ & = R^0_A(\lambda|\lambda_{0}) 
e^{f(\lambda)}.  
\end{align}
Substitution in the crossing probability yield 
\beq
\elabel{eq:crossprob}
 P_A^{MC}(\lambda|\lambda_{0})= C  \int^{\lambda}_{\lambda_n}
 R^0_A(\lambda|\lambda_{0}) e^{f(\lambda)}   d \lambda
\eeq

This is the proper reweighting of the crossing probabilities.

We are now in the position to check whether the reweighting assumption
\eref{eq:reweight} is correct, by using \eref{eq:final} and substitute
\eref{eq:crossprob}.
\begin{align}
e^{F{i} }  
 &= \frac{
   P(\lambda_{i}|\lambda_{0})  - P(\lambda_{i+1}|\lambda_{0}) } {P^0
   (\lambda_{i}|\lambda_0) -P^0 (\lambda_{i+1}|\lambda_0) } \notag \\ &
= \frac{  \int^{\lambda}_{\lambda_n}
 R^0_A(\lambda|\lambda_{0}) e^{f(\lambda)}   d \lambda -
  \int^{\lambda_{i+1}}_{\lambda_n}  R^0_A(\lambda|\lambda_{0}) e^{f(\lambda)}   d \lambda}
{ \int^{\lambda}_{\lambda_n}
 R^0_A(\lambda|\lambda_{0}) d \lambda -   \int^{\lambda_{i+1}}_{\lambda_n}
 R^0_A(\lambda|\lambda_{0})   d \lambda } \notag \\ &= \frac{  
  \int_{\lambda_{i+1}}^{\lambda_i}  R^0_A(\lambda|\lambda_{0}) e^{f(\lambda)}   d \lambda}
{  \int_{\lambda}^{\lambda_i}
 R^0_A(\lambda|\lambda_{0})   d \lambda } = e^{f(\lambda_i)},  
\end{align}
where the latter equality follows from the fact that all paths between
$\lambda_i$ and $\lambda_{i+1}$ get the same weight.

\begin{table*}[t]
  \caption{Contributions to the KL divergence of the path ensembles for the 2D potentials as function of the CV as given by \eref{eq:kldivalpha}.\label{tab:kldiv}. Due to the symmetric potential we can set $\alpha_0 = 0.5 $. }
  \begin{tabular}{llllllllllll}
    \hline
CV  & $\mu_A$ & $\mu_B$ & $S_A$ & $S_B$ & $S_A +S_B $ & $\alpha$ & correction & $\alpha S_A$   & $(1-\alpha)   S_B$  & $\alpha S_A + (1-\alpha) S_B$ &  S corrected \\ 
    \hline
   x & -1 & -1 & 0.000250302 & 0.000187073 & 0.000437375 & 0.499985 & 4.22597e-10  &0.000125147 & 9.35393e-05 & 0.000218687 & 0.000218687\\
     x+1 &-1 & -1 & 6.35347e-05 & 4.73168e-05 & 0.000110852 & 0.499984 & 5.23817e-10 & 3.17663e-05 & 2.36592e-05 &5.54255e-05 & 5.5426e-05\\
    \hline
   x  &  -1 & 0 & 0.00012251 & 0 &0.00012251 & 0.731052 & 0.110937 & 8.95608e-05 & 0 &8.95608e-05 &0.111027\\
x+y & -1 & 0 & 5.54697e-05 & 0&  5.54697e-05 &  0.731041 & 0.110927 & 4.05506e-05 & 0 & 4.05506e-05 & 0.110968\\
    \hline
x & -1 & 1 & 7.9667e-05 &  0.000489041 & 0.000568708 & 0.880779 & 0.327778 & 7.01691e-05 & 5.83039e-05 & 0.000128473 & 0.327906\\
x+y &    -1  & 1 & 6.376e-05 & 0.000163754 &0.000227514 &0.880764 & 0.327746 & 5.61575e-05 & 1.95255e-05 & 7.5683e-05 & 0.327822\\
   \hline
x & -2 & 2 & 0.000137544 & 0.00888312 & 0.00902066 & 0.982 & 0.602998  &0.000135069 & 0.000159894 & 0.000294963 & 0.603293\\
    x+y & -2 &2 & 0.000152529 & 0.00166459 & 0.00181711 & 0.981987 & 0.602943 & 0.000149781 & 2.9985e-05 & 0.000179766 &0.603123\\
    \hline
x &    -3 &3 &0.000146041 &0.0517943 &0.0519403& 0.997521 &0.675799& 0.000145679 &0.000128381& 0.00027406 &0.676074\\
x+y &-3 &3 &0.000254962 &0.0136426 &0.0138976& 0.997514 &0.675753 &0.000254328 &3.39218e-05 &0.00028825 &0.676041\\
  \hline
\end{tabular} 
\end{table*}

\section{The committor is a solution for $g(\lambda$)}
The natural solution of the equation 
\beq
\frac{\int d\lambda g(\lambda) \rho(\lambda) }{ \int d\lambda \rho(\lambda)} = K 
\eeq
is that $g(\lambda)$ is equal to the committor $p_B(\lambda)$. 
This can be seen as follows. 
The definition of the committor is  the fraction of all paths that lead to $B$
\beq
p_B(\lambda) =  \frac{\rho_{BB} (\lambda)+ \rho_{AB}  (\lambda)}{\rho (\lambda)}
= \frac{\rho_B (\lambda)}{\rho (\lambda)}
\eeq
where  $\rho_B = \rho_{BB} + \rho_{AB} $ is the density of points that commit to B.
As the total density $\rho(\lambda) = \rho_A (\lambda)+ \rho_B(\lambda)  $.
Substituting $g(\lambda)$ by $p_B(\lambda)$ gives 
\beq
\frac{\int d\lambda p_B(\lambda) \rho(\lambda) }{ \int d\lambda \rho(\lambda)} = \frac{\int d\lambda \rho_B(\lambda) }{ \int d\lambda \rho(\lambda)}, 
\eeq
which is indeed the fraction of points committed to B, thus by definition equal to $K^{exp}$. 

So the task is now to find the function $p_B(\lambda)$ that solves 
\beq
\frac{\int d\lambda \rho_B(\lambda) }{ \int d\lambda \rho(\lambda)} =K^{exp}
\eeq
given that original distributions obey
\beq
\frac{\int d\lambda \rho^0_B(\lambda) }{ \int d\lambda \rho^0(\lambda)} =K^0
\eeq
Indeed, using the ME reweighted densities this leads to the self consistent relation for the committor \eref{eq:implicitpb}.

\begin{figure}[t]
\includegraphics[width=1\linewidth]{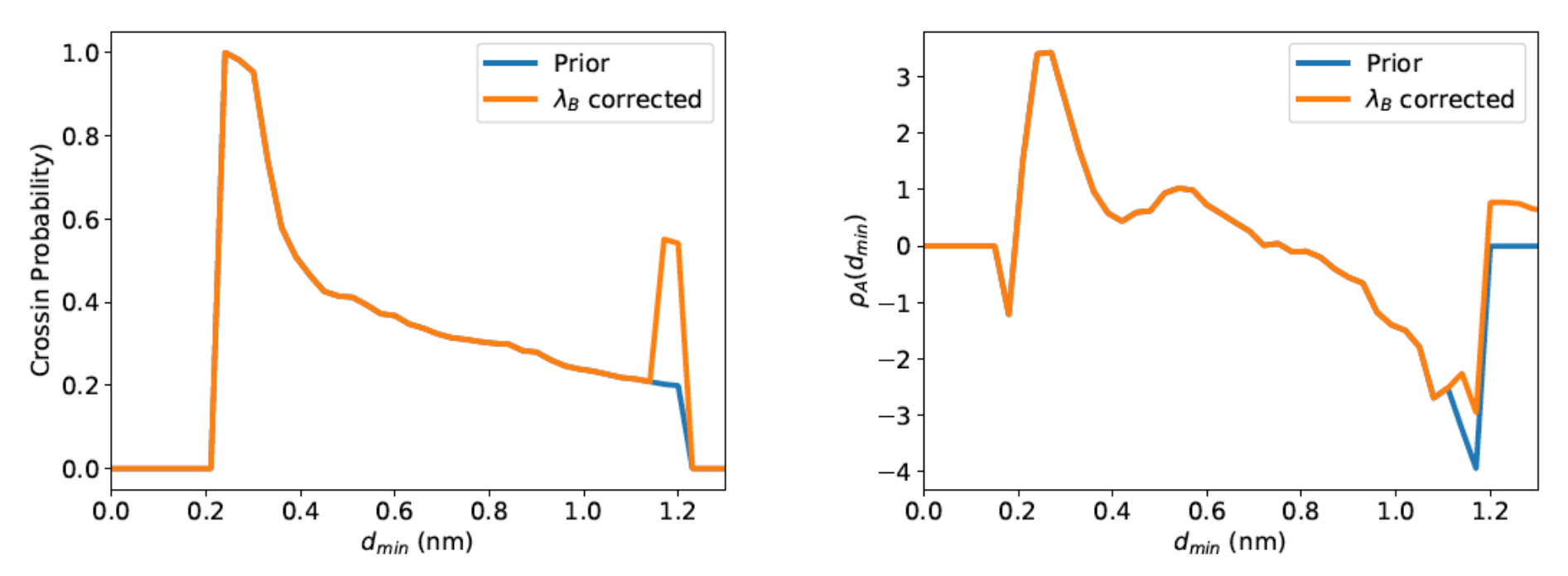}
\caption{\flabel{figAP}  a) Crossing probability out of state A and b) the respective configurational density. }
\end{figure}

\section{Restraining only the final interface $\lambda_B$} 
\label{sec:appC}

Here, we demonstrate how reweighting only pathways that cross the interface of state B leads to and  a discontinuity and sudden increase in the otherwise monotonically decreasing crossing probability histogram, (see ~\fref{figAP}a) as well as a sudden discontinuity in the configurational density ( ~\fref{figAP}b).

\section{The MaxCal entropy for different CVs in the 2D potential}
\label{sec:appkldiv}
In section  \ref{sec:optcv} we introduced  \eref{eq:kldivalpha} to compute the MaxCal entropy for the optimised path distributions.
Using this equation we can compute the entropy or equivalently the KL divergence for the 2D potentials using different bias settings for two different CVs: $\lambda=x$ along the x axis (or equivalently the y axis), and along the diagonal  $\lambda= x+y$.
In  Table \ref{tab:kldiv} we report the values for the individual path distributions $S_A$,$S_B$, as well as the total   $S_A +S_B $, using   \eref{eq:KLdivmt}. In addition, we compute the correction based on the $\alpha$ parameter due to the normalisation, from \eref{eq:kldivalpha} . The last three columns report the $\alpha$ weighted entropies (KL divergence), as well as the  corrected value of $S$. Note that we report the negative entropy, or KL divergence, in order to keep all numbers  positive. Also, here we can set $\alpha_0=0.5$ due to the symmetry of the potential.

Note that the diagonal CV is always performing better for the total path distributions ($S_A +S_B $). This is to be expected, as the diagonal is more  aligned to the reaction coordinate, and thus the distribution are expected to be less perturbed.    However, for the individual AB and BA path distribution a strong bias  (e.g. $\mu_A=-2$) will make the diagonal CV for the AB distribution seem worse (4th column), indicating that diagonal AB distribution is perturbed more than the  AB distribution for $\lambda=x$. This is possibly caused by difficulties in the numerical solution of the Volterra equation.
We stress that the MaxCal approach should focus on the total distribution, to make a proper comparison, and indeed, both the sum $S_A+ S_B$ and the alpha weighted, and total corrected entropy all show improvements for the diagonal.  

Note also that the final KL divergence reported in the last column, including the correction, is steadily increasing when the bias becomes more asymmetric. In fact it is largely dominated by the correction term, due to this asymmetry. This reflects the fact that due to the  asymmetry the BA and AB distributions are biased in different directions. However,  when just the individual distribution are considered, the normalisation reduces the individual entropies. When the distributions are taken together, the larger difference in weighting (the value of $\alpha$) is causing a much larger KL divergence.  It is thus the asymmetry, reflecting a change in the thermodynamic equilibrium constant that dominates the MaxCal entropy.
Note that for symmetric bias the correction  vanishes.

\bibliography{opslib,library}

\end{document}